\documentclass[lettersize,journal]{IEEEtran}
\usepackage{amsmath,amsfonts}
\usepackage{algorithmic}
\usepackage{algorithm}
\usepackage{array}
\usepackage[caption=false,font=normalsize,labelfont=sf,textfont=sf]{subfig}
\usepackage{textcomp}
\usepackage{stfloats}
\usepackage{url}
\usepackage{verbatim}
\usepackage{graphicx}
\usepackage{cite}
\usepackage{comment}
\usepackage{color}

\usepackage{hyperref} 

\makeatletter
\apptocmd\@maketitle{{\myfigure{}\par}}{}{}
\makeatother
\usepackage{capt-of}
\usepackage{cleveref}
\usepackage{booktabs}
\usepackage{tabularx}
\usepackage{multirow}
\usepackage{pifont}
\usepackage{textcomp}
\usepackage{xcolor}

\title{SDGraph: Multi-Level Sketch Representation Learning by Sparse-Dense Graph Architecture}

\author{
Xi Cheng$^1$, Pingfa Feng$^{1, 2}$, Mingyu Fan$^1$, Zhichao Liao$^1$, Hang Cheng$^1$, Long Zeng$^{1, \dag}$

\thanks{
This work was supported by the National Key Research and Development Program of China under Grant 2022YFB3303101.

$^1$Tsinghua Shenzhen International Graduate School, Tsinghua University, Shenzhen 518055, China.

$^2$Department of Mechanical Engineering, Tsinghua University, Beijing 100084, China.

$\dag$: Corresponding author}
}

\newcommand\myfigure{
\centering
\vspace{0.4cm}
\setcounter{figure}{0}
    \includegraphics[width=1\textwidth]{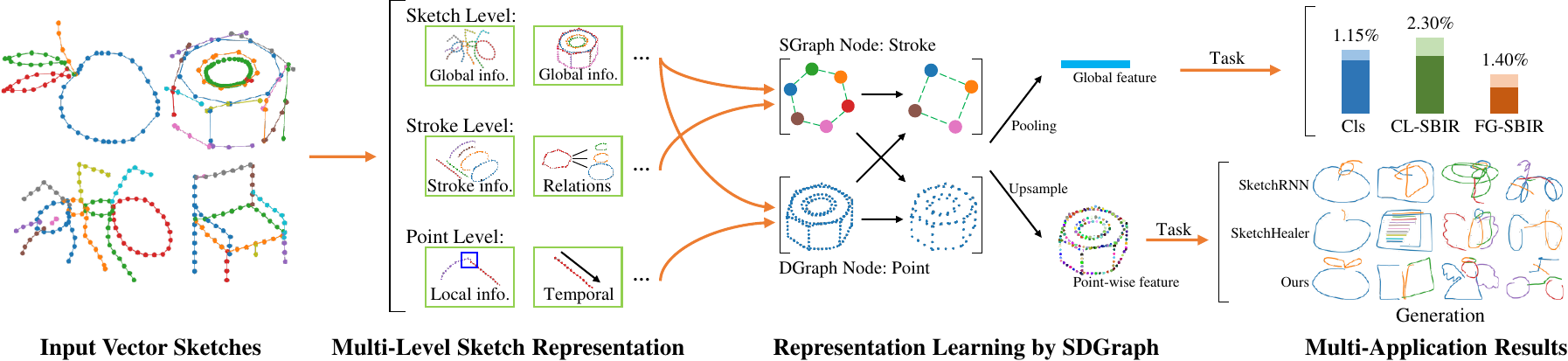}
    \vspace{-0.6cm}
    \captionof{figure}{\textbf{Method overview.} The research object of this paper is vector free-hand sketch, i.e., representing a sketch as a list of ordered points \cite{skhrnn}. We first introduced the Multi-Level Sketch Representation Scheme, to identify the information that is effective for sketch representation learning. Building upon this scheme, we proposed a deep learning architecture SDGraph, to effectively leverage all identified effective information, and output either global feature or point-wise feature to support various tasks. Extensive experiments on classification, retrieval, and generation validate the effectiveness of the proposed method.
    }
\label{fig:pilot}
\vspace{-0.4cm}
}

\begin{document}
\maketitle

\begin{abstract}

Freehand sketches exhibit unique sparsity and abstraction, necessitating learning pipelines distinct from those designed for images. For sketch learning methods, the central objective is to fully exploit the effective information embedded in sketches. However, there is limited research on what constitutes effective sketch information, which in turn constrains the performance of existing approaches. 
To tackle this issue, we first proposed the Multi-Level Sketch Representation Scheme to identify the effective information. The scheme organizes sketch representation into three levels: sketch-level, stroke-level, and point-level. This design is based on the granularity of analytical elements, from coarse (sketch-level) to fine (point-level), thereby ensuring more comprehensive coverage of the sketch information. For each level, we conducted theoretical analyses and experimental evaluations to identify and validate the effective information. 
Building on the above studies, we developed SDGraph, a deep learning architecture designed to exploit the identified effective information across the three levels. SDGraph comprises two complementary modules: a Sparse Graph that treats strokes as nodes for sketch-level and stroke-level representation learning, and a Dense Graph that treats points as nodes for sketch-level and point-level representation learning. Both modules employ graph convolution along with down-sampling and up-sampling operations, enabling them to function as both encoder and decoder. Besides that, an information fusion module bridges the two graphs to further enhance feature extraction. SDGraph supports a wide range of sketch-related downstream tasks, achieving accuracy improvements of 1.15\% and 2.30\% over the state-of-the-art in classification and retrieval, respectively, and 32.93\% improvement in vector sketch generation quality.

\end{abstract}


\section{Introduction}
\IEEEPARstart{S}{ketch} representation learning remains challenging due to the inherent sparsity and abstraction of sketches \cite{lvm, sketchy}. Although the information imbalance between sketches and images is widely recognized, to the best of our knowledge, little prior work has explored the effective information embedded in sketches. As a result, many studies tend to overlook some effective information (e.g., inter-stroke relations), thereby limiting their performance. Given the above situations, a comprehensive summary of effective sketch information, followed by the deep learning architecture that fully leverages such information, promises to improve sketch learning accuracy and broaden the applicability of sketch-based methods.

Existing sketch-specific learning methods often struggle to exploit all the available effective information. To ensure a more comprehensive investigation of the sketch information considered in existing literature, we conduct analyses at the sketch-level, stroke-level, and point-level, which are defined under our proposed Multi-Level Sketch Representation Scheme. For sketch-level representation, prior works primarily focused on global information; its effectiveness is unquestionable \cite{skh_a_net, sia_revl}, but there is still effective information at other levels to consider. For stroke-level representation, existing research utilized the intra-stroke information and inter-stroke relations; these methods are relatively rare. SketchGNN \cite{skh_gnn} extracts stroke features by applying max-pooling over stroke point features, but it struggles to capture inter-stroke relations. S3Net \cite{s3net}, SketchXAI \cite{skhxai}, and BezierSketch \cite{bezier_skh} treated stroke as independent processing units, which may lead to the omission of fine-grained local details, such as stroke intersections. For point-level representation, local information and sketch temporal information have been considered in several studies \cite{skhrnn, skhknitter}, but those models’ architecture often limits their ability to capture stroke information effectively.


To tackle the above issues, we proposed the Multi-Level Sketch Representation Scheme and designed the Sparse Graph and Dense Graph (SDGraph) deep learning architecture. The Multi-Level Sketch Representation Scheme is a structured research about the effective sketch information under our framework, which decomposed sketch representations into sketch-level, stroke-level, and point-level. Each level is studied individually to identify its effective information. This hierarchical structure is designed according to the granularity of analytical units. The coarse-grained level is the sketch-level, where the entire sketch is taken as the analytical unit. The medium-grained level is the stroke-level, motivated by the fact that a sketch is composed of multiple strokes. The fine-grained level is the point-level, since both sketches and strokes are fundamentally constructed from points.
At the sketch level, the global information is identified as effective. At the stroke level, intra-stroke information and inter-stroke relations contribute to performance. For the point-level representation, local information and stroke point adjacency are found to be effective in sketch representation learning. 
The SDGraph is designed to leverage all effective information across multi-level sketch representations. SDGraph consists of four key modules: Preprocessing module, Sparse Graph (SGraph) module, Dense Graph (DGraph) module, and Information Fusion module. The Preprocessing module standardizes input sketches and filters out noise. The SGraph treats strokes as graph nodes to learn sketch-level and stroke-level representations. The DGraph uses sketch points as graph nodes to exploit both sketch-level and point-level representations. 
Both SGraph and DGraph support graph convolution, down-sampling, and up-sampling, thereby functioning as both encoder and decoder, i.e., mapping of an entire sketch into a feature vector $\boldsymbol{f}$ for classification and retrieval, as well as the reconstruction of $\boldsymbol{f}$ into point-wise features for sketch generation. 
The Information Fusion module enables mutual exchange and integration of features between the SGraph and DGraph, enhancing the overall efficiency and completeness of information utilization.

Our contributions are threefold:
\begin{itemize}
    \item We proposed the Multi-Level Sketch Representation Scheme, which is a structured research of effective information in sketch-level, stroke-level, and point-level sketch representations under our framework, providing a solid foundation and valuable reference for future research in free-hand sketch representation learning.
    
    \item We designed the SDGraph, a deep learning architecture that leverages all the effective information across multi-level sketch representations. Moreover, SDGraph is compatible with a wide range of freehand sketch-related downstream tasks.
    
    \item We conducted extensive experiments on classification, retrieval, and generation, which validate both the effectiveness of the Multi-Level Sketch Representation Scheme and the superiority of the proposed SDGraph architecture.
    
\end{itemize}

\section{Related Work}

\textbf{Sketch-level representation learning: }
Sketch-level representation generally includes global information only. Since local information (point-level for vector sketches, or pixel-level for raster sketches) usually serves as the foundation of global information, prior works that leverage local information are also reviewed here. Local information is typically extracted by convolutional neural network (CNN) filters or point‐based neighborhood operators (e.g., K-Nearest Neighbors: KNN), while global information is generally obtained by aggregating local features (e.g., max‐pooling). The importance of local and global information is unquestionable, and nearly all sketch learning methods attempt to utilize both \cite{skh23, dood, mixsa, skhaug, bbd, skh_a_net, skhnet, skhformer, skhpnet, spfnet, dafu}.

Sketch-a-net \cite{skh_a_net} extracts local and global information based on CNN filters and max‐pooling, respectively. The input sketches are augmented by removing simple strokes and applying local distortions. These augmented sketches are then fed into CNNs to extract local and global information. SketchNet \cite{skhnet} is similar to sketch-a-net, adopts weakly supervised learning by feeding sketch-image pairs into shared CNNs, capturing local and global information that is shared between images and sketches. Sketchformer \cite{skhformer} leverages self-attention layers and max-pooling to extract local and global information. Due to the permutation-invariant nature of the attention mechanism, Sketchformer does not pay attention to the sketch’s temporal information, but allows the network to decide which timesteps to focus on. SketchPointNet \cite{skhpnet} uses KNN for local information extraction and max-pooling for global information aggregation, in which sketches are treated as 2D point clouds, and processed using a PointNet-like architecture \cite{pnet2, cstnet} for classification. Considering the permutation-invariant nature of point clouds, the sketch temporal information is deprecated. Spfusionnet \cite{spfnet} extracts local information using both CNN filters and KNN, and obtains global information through max-pooling. The model takes as input both vector sketches and their corresponding raster sketches. The vector sketches are processed by 2D PointNet++ \cite{pnet2}, while the raster sketches are fed into CNNs, enabling more effective extraction of local information by leveraging complementary representations, but also risking information redundancy.

\textbf{Stroke-level representation learning: }
Stroke-level representation includes intra-stroke information and inter-stroke relations. Intra-stroke information refers to individual stroke attributes such as stroke shape and position, while inter-stroke relations capture structural patterns between strokes, e.g., parallelism, symmetry, and spatial alignment. In practice, intra-stroke information is typically extracted by feeding either an image or a point set of an individual stroke into neural networks. Inter-stroke relations are usually captured by further processing the extracted intra-stroke features through relational or graph-based networks \cite{skhjew, skhpp, freegen, clipasso, clipascene, skhpq, bezier_skh, s3net, skh_gnn, dgcnn, skhxai, sketch_gloc}.

Few existing studies explicitly consider stroke-level information. BezierSketch \cite{bezier_skh} extracts intra-stroke information by feeding stroke points into bidirectional RNNs, and treats each stroke as a separate processing unit to achieve inter-stroke relations extraction. However, this approach struggles to capture fine-grained local features between strokes, such as stroke intersections and stroke end snapping. S3NET \cite{s3net} emphasizes that the mid‐length strokes are more suitable for sketch structural modeling. To this end, it splits long strokes into mid-length segments, and represents the sketch as a graph of these stroke segments. Graph Neural Networks (GNNs) are then employed to extract intra-stroke information and inter-stroke relations from these segments. Nevertheless, as S3NET treats segments as independent units, it remains limited in capturing fine-grained local information between strokes. SketchGNN \cite{skh_gnn} constructs a graph over sketch points, where edges connect adjacent points within the same stroke. This graph is processed by DGCNN \cite{dgcnn} to extract point-wise features, and intra-stroke information is subsequently obtained by aggregating these point features for each stroke. However, due to the absence of inter-stroke edges in the graph, this method is limited in its ability to capture inter-stroke relations. SketchXAI \cite{skhxai} decomposes stroke information into three components: drawing order, stroke shape, and stroke position. These components are separately encoded to extract intra-stroke information, and the encoded features are then fed into bidirectional LSTMs \cite{lstm} to capture inter-stroke relations. Nonetheless, since this method processes strokes individually, it struggles to capture fine-grained local information between strokes.

\textbf{Point-level representation learning: }
Point-level representation includes local information, temporal information, and point sampling frequency. Temporal information refers to the sequential order in which sketch points are drawn; it is an inherent and distinctive property of vector sketches that is absent in raster sketches. Many existing methods employ recurrent architectures to extract temporal information. The point sampling frequency is reflected in the varying spatial distances between consecutive points, which implicitly encode stroke speed and drawing dynamics. Nearly all methods that operate on vector sketches implicitly leverage this information during processing \cite{fan, skhrnn, attn_net, empri, mgt, skhbert, skhr2cnn, skhmate, aiskher, deepskh2, rfp}.

SketchRNN \cite{skhrnn} extracts temporal information using a bidirectional LSTM \cite{lstm}, and incorporates a variational autoencoder \cite{vae} to construct a continuous latent space, enabling both sketch recognition and generation. While SketchRNN popularized vector sketch modeling and enabled controllable sketch synthesis, its reliance on LSTM results in generated outputs often suffer from mode collapse or fine-grained detail lost. Attention-Net \cite{attn_net} uses temporal convolutional networks (TCNs) to extract temporal information, which has been shown to outperform RNNs in sequence modeling tasks \cite{empri}. MGT \cite{mgt} extracts temporal information by directly encoding the point indices of the sketch point sequence. Each sketch point is expressed as $[(x,y),s,i]$, where $(x,y)$ is the coordinate, $s$ is the stroke‐end flag, and $i$ is the sequence index. These triplets are processed by multi‐head attention to capture temporal information. MGT narrows the performance gap between vector-based and raster-based sketch learning methods, and outperforms RNN-based approaches on classification tasks. SketchBERT \cite{skhbert} extracts temporal information by embedding the point’s sequence index. It employs a novel “Sketch Gestalt Model” for pre-training, and is subsequently fine-tuned on recognition and retrieval tasks, achieving superior performance over generative pre-training approaches. Sketch-R2CNN \cite{skhr2cnn}, SketchMate \cite{skhmate}, and AI-Sketcher \cite{aiskher} extract temporal information by RNN or bidirectional LSTM. These methods adopt a hybrid architecture that combines an RNN branch for processing vector sketches, and a CNN branch for processing rasterized sketches. While this design enhances sketch representation learning, it also leads to increased latency and memory consumption. DeepSketch 2 \cite{deepskh2} extracts stroke temporal information by constructing a rasterized cumulative stroke‐sequence $\mathcal{S}=\left\{\boldsymbol{I}_1,\boldsymbol{I}_2,\dots,\boldsymbol{I}_N\right\}$, where $\boldsymbol{I}_k$ represents the sketch after the k-th stroke is drawn. Each frame $\boldsymbol{I}_k$ is processed by CNNs to extract a corresponding feature $\boldsymbol{f}_k$, and the sequence $\left\{\boldsymbol{f}_1,\boldsymbol{f}_2,\dots,\boldsymbol{f}_N\right\}$ is processed by LSTM to model temporal stroke evolution.

Although existing methods leverage a wide range of sketch information, to the best of our knowledge, little prior work has systematically investigated which types of information are effective for sketch representation learning, nor have they fully exploited all available effective information. Addressing these gaps is crucial for improving the accuracy of sketch learning methods and enhancing their applicability in real-world scenarios.

\section{Method}
\subsection{Multi-Level Sketch Representation Scheme}

The Multi-Level Sketch Representation Scheme is developed to identify the effective information embedded in sketches. It consists of a hierarchical sketch representation structure, and a summary of the effective information within each level. The hierarchical structure is designed to study sketch information as comprehensively as possible. For each level, we identify the effective information by theoretical analysis and experimental validation.

The hierarchical structure includes sketch-level, stroke-level, and point-level representations. This design is motivated by the intrinsic multi-granularity nature of sketches, covering information from the overall composition, to mid-level stroke structures, and fine-grained point patterns. Such a hierarchy mirrors the progressive process of human sketching, from global planning to structural construction to fine execution, while enabling the use of scale-specific learning mechanisms (e.g., global feature aggregation, graph-based stroke modeling, and fine-grained trajectory encoding).

The analysis and summary of the effective information at each level are as follows:

\textbf{Sketch-Level Representation:} Taking the entire sketch as the basic analytical unit, describing its global structure.

Sketch-level representation contains the global information, as it is derived by taking the entire sketch as the analysis unit. global information captures the sketch’s overall structure, and its effectiveness is widely acknowledged.

\textbf{Stroke-Level Representation:} Taking individual strokes as the basic analytical unit, characterizing their geometric properties, internal structure, and relationships with other strokes.


By treating individual strokes as the analysis units, we define stroke-level representation to include intra-stroke information, inter-stroke relations, inter-stroke temporal information, and intra-stroke drawing direction. Through our analysis and experiments, we find that the effective information consists of intra-stroke information and inter-stroke relations. Inter-stroke temporal information and intra-stroke drawing direction does not contribute to the performance under our settings, and is therefore not considered, but it may be effective for other methods. 


Intra-stroke information characterizes individual strokes in terms of properties such as length, curvature, and shape. In freehand sketches, each stroke often conveys explicit and intentional design cues. For example, a circular stroke may signify a hole in a mechanical part.

Inter-stroke relations refer to the spatial and geometric relationships between different strokes, such as proximity, overlap, parallelism, and symmetry. Inter-Stroke Relations are crucial for inferring the sketcher's intent. For instance, when drawing a rectangle, the sketcher typically maintains parallelism between opposite sides and perpendicularity between adjacent sides.


Inter-stroke temporal information reflects the drawing order of strokes. Prior works \cite{sketchhealer, dcgra} have demonstrated that stroke order is beneficial for autoregressive generation frameworks, where sketches are produced point-by-point using LSTM-based decoders. In such settings, the stroke order naturally serves as an explicit supervisory signal. In contrast, this paper adopts a diffusion-based generative paradigm \cite{ddpm}, which generates the entire sketch in a parallel manner, i.e., all strokes are generated simultaneously. Consequently, stroke order is not an inherent component of our generation process. Moreover, stroke order does not affect the sketch’s visual appearance, and exhibits substantial variability across users \cite{free2cad, attn_net}. Therefore, we do not consider the inter-stroke temporal information under our specific settings.


Stroke drawing direction does not affect the appearance of the sketch, and the drawing direction of the same stroke may differ for different artists; therefore, it is not considered.

\textbf{Point-Level Representation:} Taking discrete points along sketched as the basic analytical unit, describing local regions features, as well as temporal and frequency-related characteristics.

By treating individual points as the analysis units, we define point-level representation to local information, stroke point adjacency, and point frequency. Through our analysis and experiments, we identify local information and stroke point adjacency as effective. Point frequency information does not contribute to the performance under our settings, and is therefore not considered.

Local information includes fine-grained geometric cues such as endpoint snapping, local stroke intersections, and other nearby structural features. The effectiveness of local information is unquestionable.

\begin{figure*}[!t]
\centering
\includegraphics[width=1.0\linewidth]{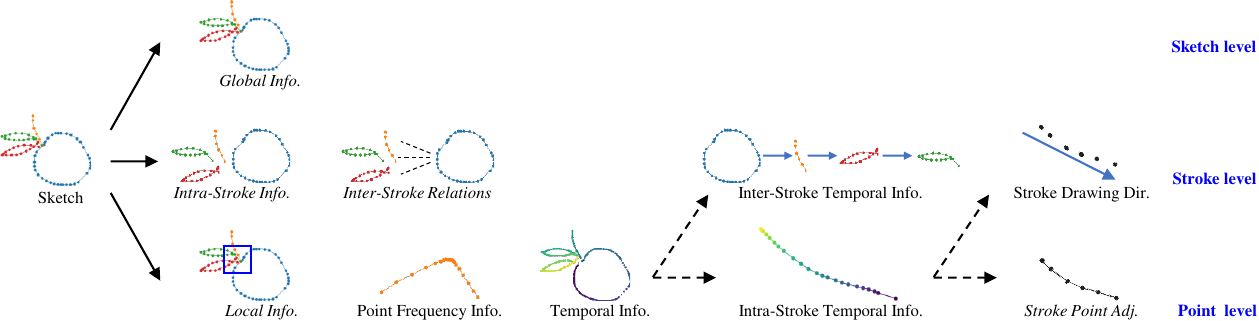}
\caption{\textbf{Multi-Level Sketch Representation Scheme.} Italic labels denote the information utilized in the proposed framework. 
Sketch temporal information is decomposed into intra-stroke and inter-stroke components. 
The intra-stroke temporal information is further characterized by the stroke drawing direction and stroke point adjacency. 
The stroke drawing direction can be represented by the stroke starting point, 
while stroke point adjacency describes the adjacency relationships between points during the drawing process.}
\label{fig:ml_frwk}
\end{figure*}

Stroke point adjacency captures the point adjacency relationships during the stroke drawing process. Compared with the intra-stroke temporal information, stroke point adjacency do not consider the absolute drawing direction; that is, for a stroke defined as $\left\{\boldsymbol{p}_i|i=1,2,\dots,N\right\}$, drawing it along $\boldsymbol{p}_1\rightarrow \boldsymbol{p}_2\rightarrow\dots\rightarrow \boldsymbol{p}_N$ is treated as equivalent to drawing it in the reverse order $\boldsymbol{p}_N\rightarrow \boldsymbol{p}_{N-1}\rightarrow\dots\rightarrow \boldsymbol{p}_1$, as shown in \Cref{fig:ml_frwk}. The point adjacency within a stroke can influence its visual appearance; therefore, it is considered effective for sketch representation learning.

Point frequency information reflects the density of sketch points. As it does not influence the sketch’s visual appearance, and varies between users, it may mislead deep learning models.

Based on our studies and existing literature, we identified the effective sketch information as illustrated in \Cref{fig:ml_frwk}, and validated its effectiveness through experiments, as presented in \Cref{tab:eff_info} and \Cref{tab:ist_info}. 

In \Cref{tab:eff_info}, each row either excludes one type of effective information or includes one type of not considered information to assess its impact. 
The SDGraph (\Cref{workflow}) and QuickDraw subset \cite{mgt} are used for experiments.
The Baseline includes all identified effective information while excluding all not considered information. 
We manipulate the inclusion or exclusion of specific information as follows: Excluding intra-stroke information: The SGraph module is removed. Excluding inter-stroke relations: Inter-stroke connections are disabled during SGraph updates. Excluding intra-stroke point adjacency: The ordering of points within each stroke is randomly shuffled. Including inter-stroke temporal information: During SGraph updates, neighboring nodes are searched according to the stroke drawing order, and an additional LSTM is incorporated to model inter-stroke temporal dependencies. Including inter-stroke drawing direction: The starting point of each stroke is duplicated to encode drawing direction. Including point frequency information: Relative point frequencies are preserved during resampling.
From the results in \Cref{tab:eff_info}, the classification accuracy consistently decreases when excluding any type of effective information or including not considered information, thus confirming that the effectiveness of each information aligns with our analysis. 

To further validate the effectiveness of inter-stroke temporal information and stroke drawing direction, we shuffle the stroke order or reverse the stroke drawing direction and feed the modified sketches into four existing methods. As shown in \Cref{tab:ist_info}, randomly shuffling the stroke order or stroke drawing direction has a negligible impact on model performance. If these types of information were truly effective, such perturbations would lead to a significant performance drop. Therefore, these results support our analysis regarding the limited effectiveness of inter-stroke temporal information and stroke drawing direction.

\begin{table}[b]
\caption{Validation experiments of effective and not considered information.}
\centering
\begin{tabularx}{1.0\linewidth}{
    >{\raggedright\arraybackslash}m{0.45\linewidth}  
    >{\raggedright\arraybackslash}m{0.28\linewidth}  
    >{\raggedright\arraybackslash}m{0.2\linewidth}  
}

\toprule
\textbf{Information} & \textbf{Process} & \textbf{Accuracy} \\
\midrule
Intra-Stroke Info.     & Exclude & 0.6497 \\
Inter-Stroke Relations & Exclude & 0.7093 \\
Stroke Point Adjacency  & Exclude & 0.7249 \\
\midrule
Inter-Stroke Temporal Info.  & Include & 0.7470 \\
Stroke Drawing Dir.  & Include & 0.7501 \\
Point Frequency Info.  & Include & 0.7335 \\
\midrule
Baseline               &         & \textbf{0.7537} \\
\bottomrule
\end{tabularx}
\label{tab:eff_info}
\end{table}

\begin{table}[htbp]
\caption{Validation experiments of inter-stroke temporal information and intra-stroke drawing direction.}
\centering
\begin{tabularx}{1.0\linewidth}{
    >{\raggedright\arraybackslash}m{0.40\linewidth}  
    >{\raggedright\arraybackslash}m{0.33\linewidth}  
    >{\raggedright\arraybackslash}m{0.25\linewidth}  
}
\toprule
\textbf{Method} & \textbf{Process} & \textbf{Accuracy} \\
\midrule
\multirow{2}*{Bi-directional GRU \cite{gru}} & None & 0.6768 \\
~ & Shuffle Strokes & 0.6881 \\
~ & Shuffle Drawing Dir. & 0.6759 \\
\midrule
\multirow{2}*{SketchRNN \cite{skhrnn}} & None & 0.6665 \\
~ & Shuffle Strokes & 0.6714 \\
~ & Shuffle Drawing Dir. & 0.6631 \\
\midrule
\multirow{2}*{MGT (Large) \cite{mgt}} & None & 0.7280 \\
~ & Shuffle Strokes & 0.7273 \\
~ & Shuffle Drawing Dir. & 0.7291 \\
\midrule
\multirow{2}*{SketchTransformer \cite{skhtrans}} & None & 0.6829 \\
~ & Shuffle Strokes & 0.6802 \\
~ & Shuffle Drawing Dir. & 0.6795 \\
\bottomrule
\end{tabularx}
\label{tab:ist_info}
\end{table}

Although both \Cref{tab:eff_info} and \Cref{tab:ist_info} are conducted on classification tasks, the derived conclusions can be transferred to both retrieval and generation under our specific settings.
For retrieval, both classification and retrieval aim to learn discriminative global sketch representations, and in our implementation, they share the same SDEncoder, as shown in \Cref{fig:clsrevl}. Consequently, the types of information that are beneficial or redundant for classification directly affect retrieval performance in the same manner.
For generation, we adopt a diffusion-based framework, which generates the entire sketch in a parallel manner, i.e., all strokes are generated simultaneously. Unlike autoregressive generation methods (e.g., LSTM-based decoders), which produce sketches point-by-point. Consequently, drawing order is not an inherent component of our generation process. Moreover, in the generation pipeline, the SDDecoder approximately inverts the SDEncoder, as shown in \Cref{fig:diff}, and the features learned by SDEncoder are also the basis for estimating noise in the diffusion-based generation process. The above supports the applicability of our conclusions to the generation task. 
We emphasize that this transferability holds under our specific SDGraph and DDPM settings, and may not directly apply to other configurations, such as LSTM-decoder-based generation frameworks.

\subsection{SDGraph Architecture}

\begin{figure*}[htbp]
\centering
\includegraphics[width=1.0\linewidth]{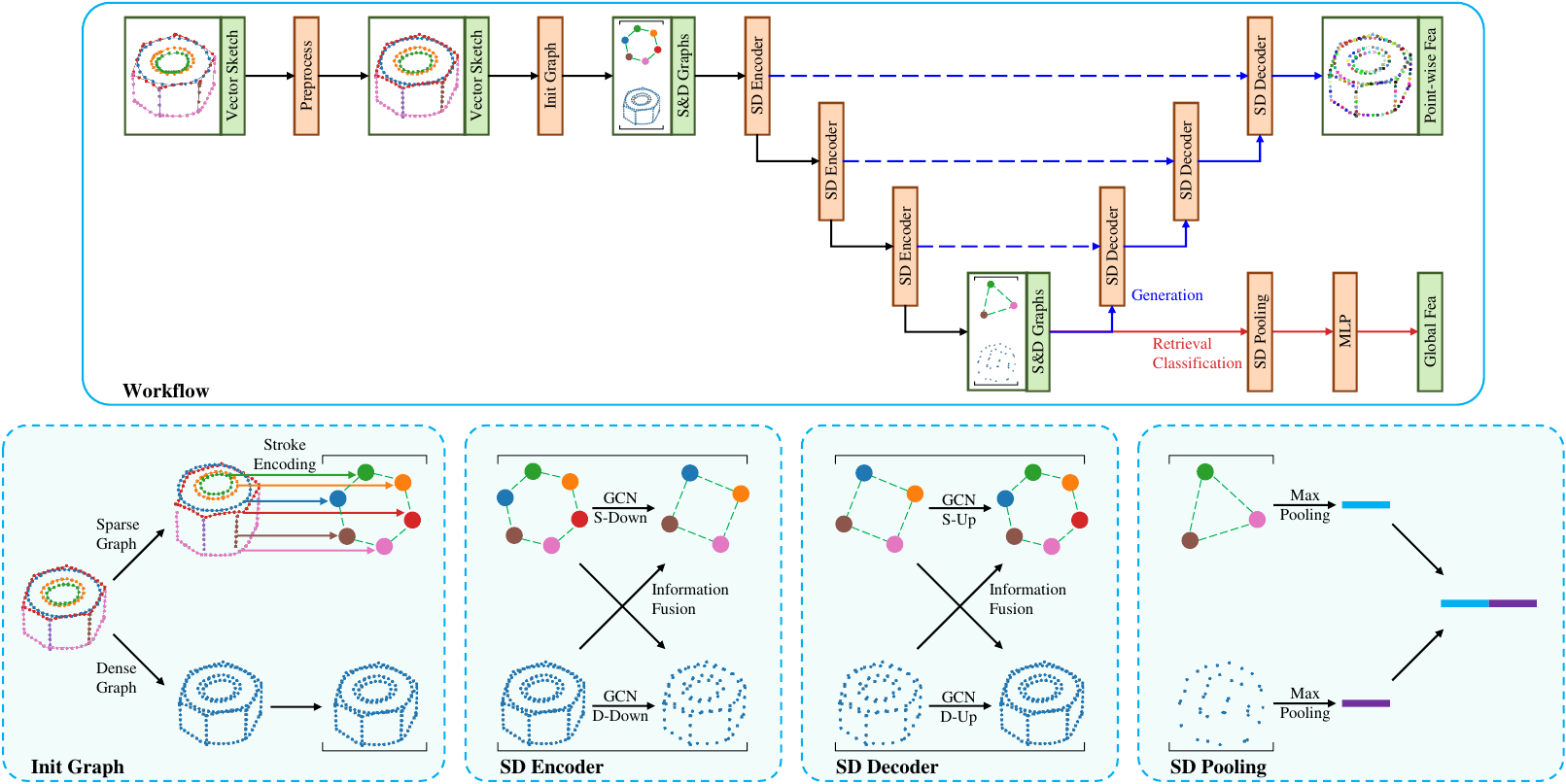}
\caption{\textbf{Overall workflow.} The input freehand sketch is first processed by the preprocess module, which filters out outliers, applies translation, scaling, and resampling to normalize the sketch. Subsequently, the init graph procedure constructs the initial SGraph and DGraph (S\&D Graphs) from the preprocessed sketch. The S\&D Graphs are then fed into the SDEncoder, which updates node features and applies down-sampling operations to accelerate feature extraction. After passing through multiple SDEncoders, the model branches based on the downstream task: for classification or retrieval, the resulting S\&D Graphs are processed by the SDPooling and Multi-Layer Perceptions (MLPs) to generate the global feature. For generation, the SDDecoder updates the node features and performs up-sampling to restore the S\&D Graphs to their original scale. Finally, SGraph features are transferred to the DGraph to produce point-wise features.}
\label{workflow}
\end{figure*}

SDGraph comprises four main modules: the preprocess module, the SGraph module, the DGraph module, and the information fusion module, as illustrated in \Cref{workflow}.

The preprocess module filters out outliers, and normalizes the point distribution through translation, scaling, and resampling. During resampling, the distances between two adjacent points are made uniform, which removes the point frequency information.

The SGraph module is designed to learn sketch-level and stroke-level representations. In this module, each node represents a stroke (group) in the sketch. The initial node features are extracted from individual stroke points, and then fed into Graph Convolutional Networks (GCNs) to capture intra-stroke information and inter-stroke relations. Down-sampling or up-sampling operations are applied as necessary. Global information is obtained by applying max-pooling over the node features. Since the GCN updates are permutation-invariant with respect to node order, the inter-stroke temporal information is inherently excluded.

The DGraph module is designed to learn sketch-level and point-level representations. In this module, each node corresponds to a sketch point (group). The initial node features are defined by the point coordinates. These features are subsequently updated through GCNs, followed by down-sampling and up-sampling as required. In the DGraph, The KNN operation is employed to capture local information, and the global information is obtained by max-pooling. Down-sampling and up-sampling are implemented via convolution and transposed convolution, respectively, both of which incorporate intra-stroke adjacency information.

The Information Fusion module facilitates data exchange between the SGraph and DGraph, thereby enhancing the efficiency and effectiveness of information extraction across multi-level representations.

\subsection{Preprocess}
The preprocess module standardizes the original vector sketch into a unified format suitable for deep learning. It consists of the following steps:
\begin{enumerate}
    \item Shift the sketch’s mass center to the origin $(0, 0)$.
    \item Normalize the sketch by scaling its bounding box to the range $[-1, 1]$.
    \item Remove outliers.
    \item Remove strokes containing too few points.
    \item Remove strokes with excessively short lengths.
    \item Resample sketch points at fixed spatial intervals to ensure uniform point distribution.
\end{enumerate}

\begin{figure}[htbp]
\centering
\includegraphics[width=1.0\linewidth]{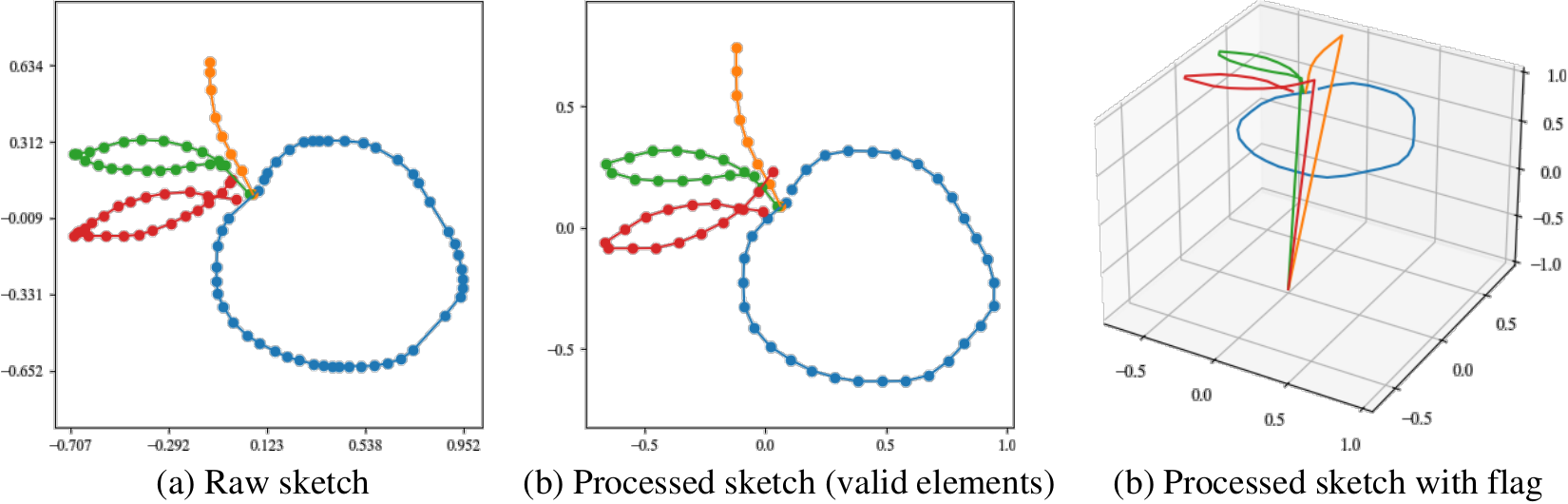}
\caption{\textbf{Raw and processed sketches.} 
After preprocessing, the sketch is normalized by aligning its centroid, scaling it with respect to the bounding box, and unifying the distances between adjacent points. 
Valid positions are represented as $(x, y, 1)$, while padded positions are encoded as $(0, 0, -1)$, as illustrated in the rightmost figure.}
\label{fig:preprocess}
\end{figure}

To enable batch processing for sketches with variable numbers of strokes and points, 
we adopt a padding strategy. Specifically, each point is represented as a 
three-dimensional feature vector $(x, y, flag)$, where $(x, y)$ denotes the normalized 
coordinates, and $flag$ indicates the validity of the point. For valid points we set 
$flag=1$, while padded points are represented as $(0, 0, -1)$. Since all coordinates 
are normalized to the range $[-1, 1]$, the additional validity indicator allows the 
network to distinguish padded elements from real sketch points. The raw sketch and its processed representation are shown in \Cref{fig:preprocess}.

For the classification and retrieval tasks, a masking mechanism is further applied 
so that padded elements do not contribute to feature aggregation during graph 
convolution and pooling operations.
In cases where a sketch contains extremely few strokes, which may cause instability 
for neighborhood-based graph operations, we apply a cycle repeat strategy to duplicate 
existing strokes until the minimum stroke requirement is satisfied.

For the sketch generation experiments, the padded representation is directly used 
as the model input, the network directly predicts the padding $flag$, and padded points are filtered during post-processing to recover the final sketch.

\subsection{Sparse Graph}
SGraph includes four submodules: 
\begin{enumerate}
    \item \textbf{Stroke encoding:} Extracts individual stroke features from the stroke points.
    \item \textbf{Graph node feature update:} Updates SGraph node features.
    \item \textbf{SGraph down-sampling (S-Down):} Reduces the number of graph nodes to accelerate computation and aggregate contextual information.
    \item \textbf{SGraph up-sampling (S-Up):} Recovers node resolution for detailed feature restoration.
\end{enumerate}

\begin{figure}[htbp]
\centering
\includegraphics[width=0.80\linewidth]{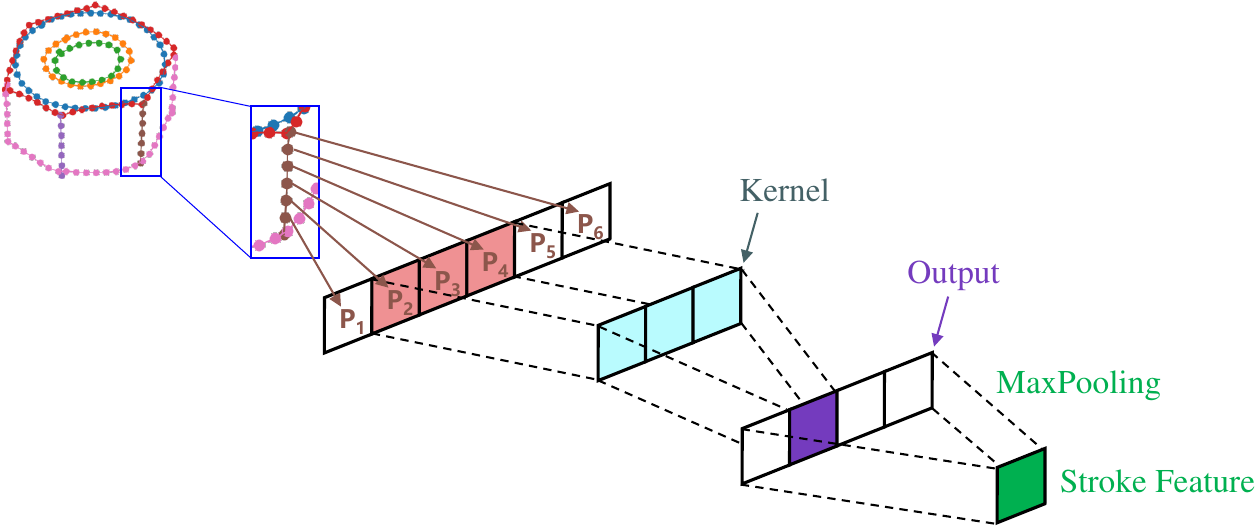}
\caption{\textbf{Stroke encoding module.} The points of each individual stroke are processed by the 1D convolution layers, and the resulting features are aggregated via max-pooling to obtain the stroke representation.}
\label{fig:stkenc}
\end{figure}

\textbf{Stroke encoding:}
The stroke encoding module is applied to the point sequence of each stroke independently, to extract initial stroke-wise features. These features are unaffected by other strokes. As illustrated in \Cref{fig:stkenc}, the feature extraction process employs a combination of convolutional layers and max-pooling operations, effectively encoding the geometric and sequential characteristics of individual strokes.

\begin{figure*}[htbp]
\centering
\includegraphics[width=0.8\linewidth]{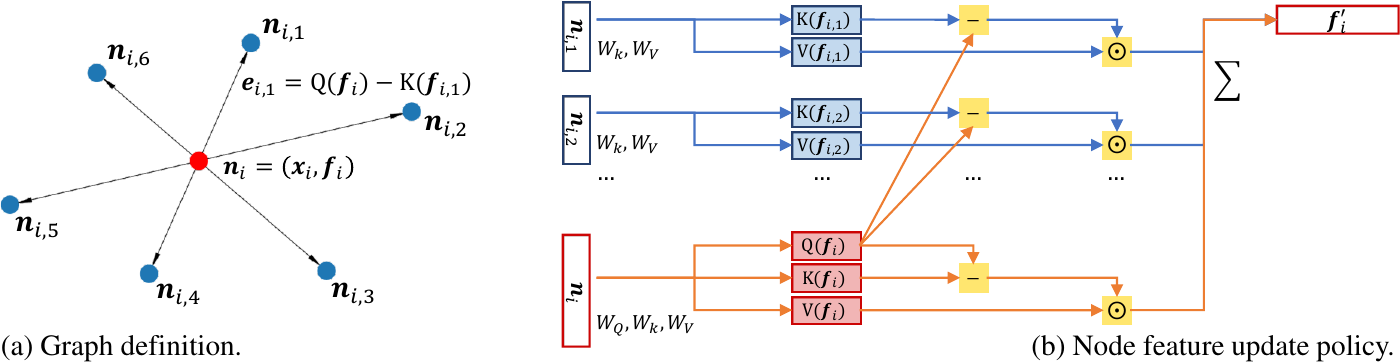}
\caption{\textbf{Graph Convolutional Network structure.} (a) A node $\boldsymbol{n}_i$ include coordinate $\boldsymbol{x}_i$ and feature $\boldsymbol{f}_i$. For SGraph, $\boldsymbol{x}_i$ is defined as the SGraph embedding generated by the Init Graph module (left bottom of \Cref{workflow}), and the number of neighbors is 2. For DGraph, $\boldsymbol{x}_i$ is defined as the point coordinate, and the number of neighbors is 30. Each node is connected with its neighbor nodes; the neighbor nodes are searched by the KNN. The edge feature is defined as 
$\boldsymbol{e}_{i,j}=\mathrm{Q}(\boldsymbol{f}_i) - \mathrm{K}(\boldsymbol{f}_{i,j})$. (b) The node feature $\boldsymbol{f}_i$ is updated by vector attention mechanism \cite{vec_attn}, the attention weight is defined by the graph edge feature $\boldsymbol{e}_{i,j}$, and the sum of all weighted node features is the updated node feature $\boldsymbol{f}_i'$.}
\label{fig:gcn}
\end{figure*}

\textbf{Graph Node feature update:}
This module updates the SGraph node features using GCNs, specifically employing GCNs \cite{dgcnn} with vector attention mechanism \cite{vec_attn} in this paper, which updates the node features by aggregating the weighted edge features connecting the node, as shown in \Cref{fig:gcn} and \Cref{eq:gcn}.

\begin{equation}
\begin{aligned}
\boldsymbol{f}_i' &= \sum_{\boldsymbol{n}_{i,j} \in \mathcal{N}_i} \rho(\boldsymbol{e}_{i,j} + \boldsymbol{\delta})\odot (\mathrm{V}(\boldsymbol{f}_{i,j}) + \boldsymbol{\delta}) \\
\boldsymbol{e}_{i,j} &= \mathrm{Q}(\boldsymbol{f}_i) - \mathrm{K}(\boldsymbol{f}_{i,j}) \\
\boldsymbol{\delta} &= \mathrm{MLP}(\boldsymbol{x}_j - \boldsymbol{x}_i)
\end{aligned}
\label{eq:gcn}
\end{equation}

Where $\boldsymbol{f}_{\mathit{i}}$ and $\boldsymbol{x}_i$ are the feature and coordinate of node $\boldsymbol{n}_{i}$, respectively. $\mathcal{N}_i$ represents all neighbor nodes around $\boldsymbol{n}_{i}$, $\rho$ is the normalization function (MLPs following softmax in this paper), and $\odot$ denotes element-wise multiplication. $\delta$ is positional encoding, $\mathrm{Q}$, $\mathrm{K}$, and $\mathrm{V}$ are MLPs.

As GCNs and attention mechanism are permutation-invariant to node order, the inter-stroke temporal information is inherently excluded during the feature updating process.

\begin{figure*}[htbp]
\centering
\includegraphics[width=0.75\linewidth]{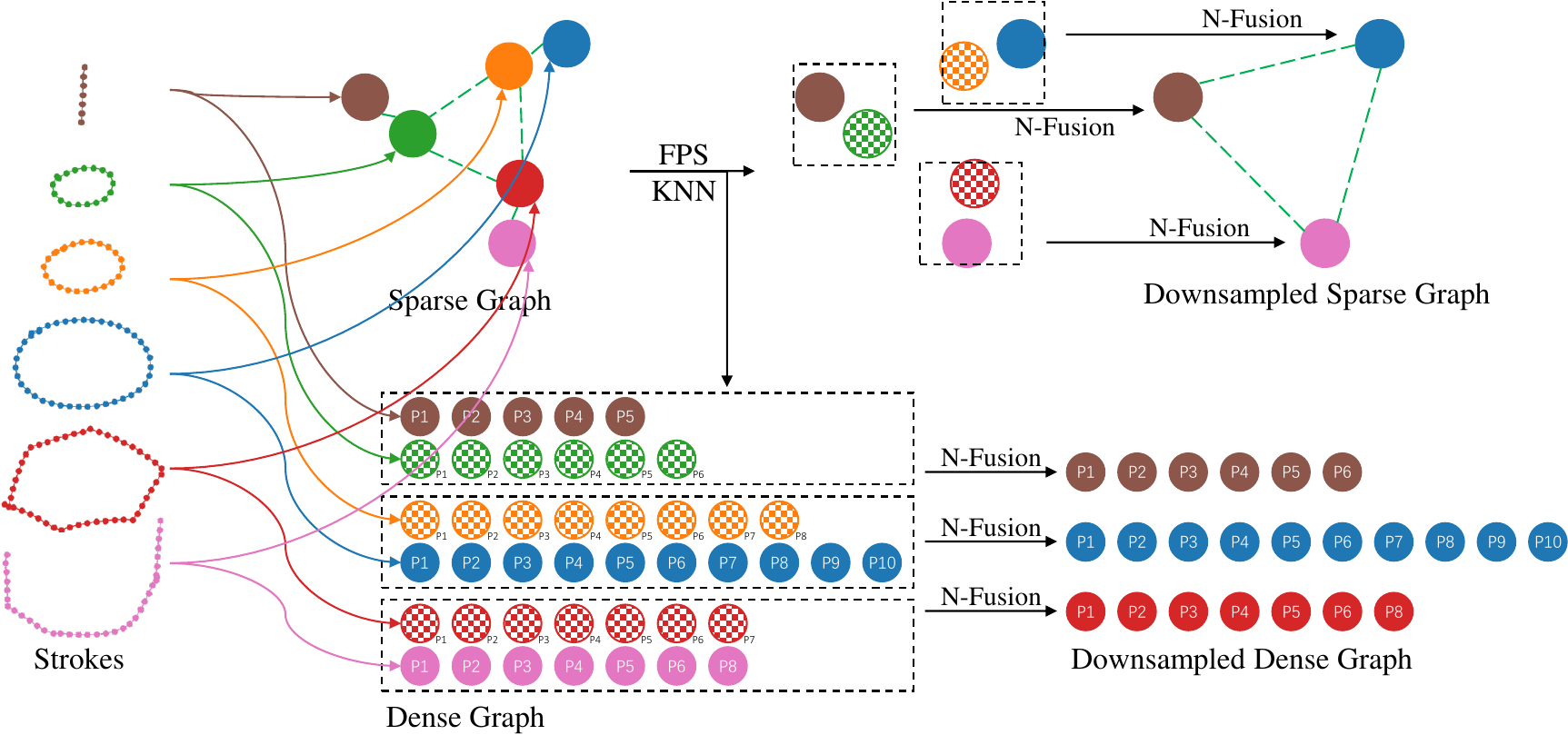}
\caption{\textbf{Sparse Graph down-sample.} \textit{Top side}: Using FPS to select a set of representative SGraph center nodes, where the initial SGraph node features serve as the stroke coordinates. Subsequently, KNN is employed to identify the neighboring stroke nodes ($k=1$ in this paper). Finally, the neighbor fusion process (N-Fusion) is performed to aggregate the neighbor node features, as shown in \Cref{fig:nfusion}. \textit{Bottom side}: The FPS and KNN results computed in the SGraph are propagated to the DGraph, and the DGraph undergoes corresponding down-sampling and N-Fusion operations.}
\label{fig:sdown}
\end{figure*}

\textbf{S-Down and S-Up:}
The S-Down module is designed to enable more efficient stroke-level feature extraction, particularly for sketches containing a large number of strokes. It is important to note that both the S-Down and S-Up operations influence the structure of the DGraph, with details in \Cref{fig:sdown}. The S-Up module employs an interpolation-based up-sampling strategy similar to that used in PointNet++ \cite{pnet2}. Simultaneously, the up-sampling parameters from the SGraph are shared with the DGraph, enabling a corresponding up-sampling operation to be performed on the DGraph.

\begin{figure*}[htbp]
\centering
\includegraphics[width=0.9\linewidth]{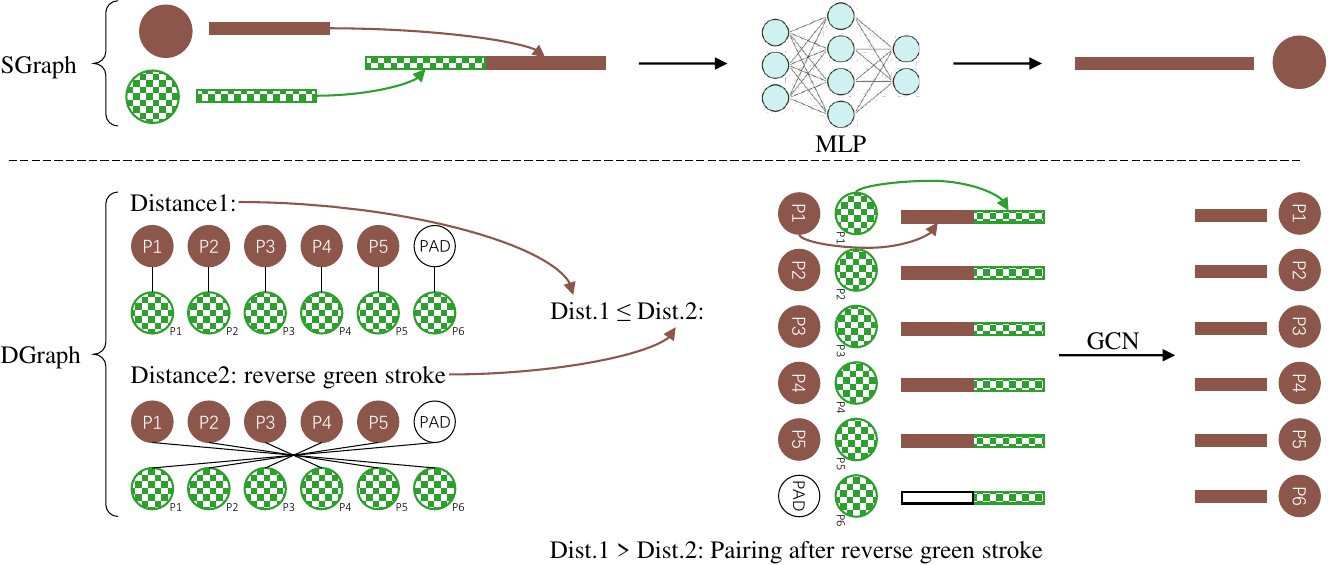}
\caption{\textbf{N-Fusion process.} \textit{Top side}: For the SGraph nodes, the features of the center nodes (brown) are concatenated with those of the neighboring nodes (green), and the concatenated feature is fed into MLPs for further processing. \textit{Bottom side}: For the DGraph nodes, compute the matching distance in both the original order and the reversed order. Specifically, given two stroke-point sequences $\{\textbf{p}_i\}_{i=0}^{n}$ and $\{\textbf{p}'_i\}_{i=0}^{n}$,
computing $\text{Dist.1} = \sum_{i=0}^{n} \| \mathbf{p}_i - \mathbf{p}'_i \|$ and $\text{Dist.2} = \sum_{i=0}^{n} \| \mathbf{p}_i - \mathbf{p}'_{n-i} \|.$ The correspondence that yields the smaller distance is selected as the final stroke–point alignment. Finally, the corresponding features are concatenated and updated via a GCN.}
\label{fig:nfusion}
\end{figure*}

\subsection{Dense Graph}
Dense Graph consists of two submodules: 
\begin{enumerate}
    \item \textbf{Graph node feature update:} Updates the features of DGraph nodes.
    \item \textbf{DGraph down-sampling (D-Down):} Reduces the number of dense graph nodes to enable more efficient feature extraction.
    \item \textbf{DGraph up-sampling (D-Up):} Increases the number of dense graph nodes to recover fine-grained features.
\end{enumerate}

\textbf{Graph node feature update:}
The GCNs shown in \Cref{fig:gcn} is employed to update the DGraph node features, where the KNN algorithm is used to capture local information.

\begin{figure}[htbp]
\centering
\includegraphics[width=1.0\linewidth]{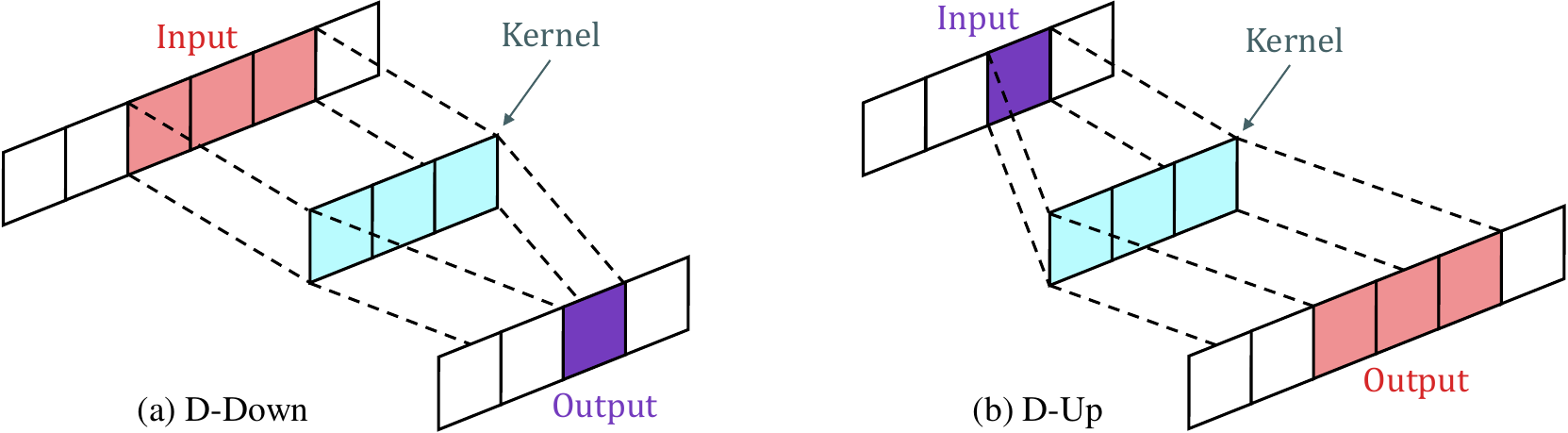}
\caption{\textbf{Sample modules in DGraph.} D-Down is implemented by 1D convolutional layers, and D-Up employs 1D transpose convolutional layers.}
\label{fig:ddownup}
\end{figure}

\begin{figure}[htbp]
\centering
\includegraphics[width=1.0\linewidth]{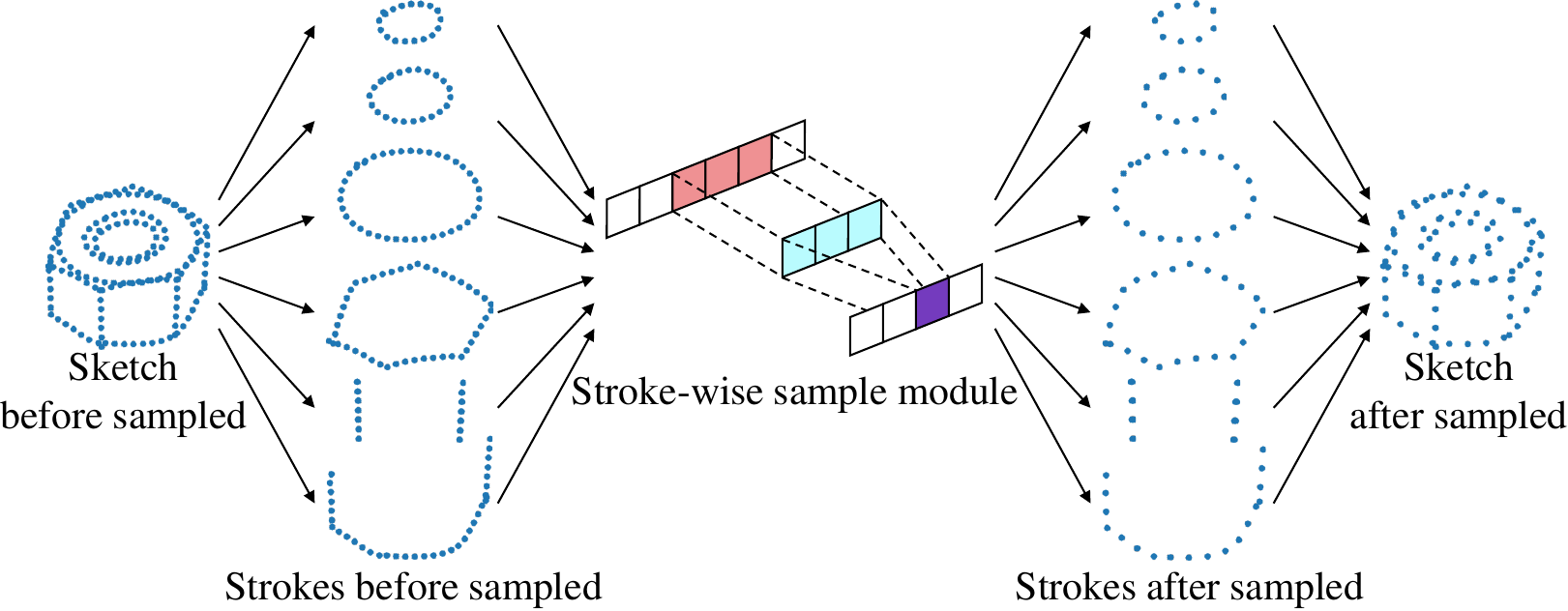}
\caption{\textbf{Stroke-wise sample process.} Points from individual strokes are sampled independently, and the resulting sampled points are subsequently merged to construct the sampled DGraph.}
\label{swsample}
\end{figure}

\textbf{D-Down and D-Up:}
The D-Down and D-Up operations are implemented using convolution and transpose convolution layers, as illustrated in \Cref{fig:ddownup}. Both down-sampling and up-sampling are performed at the stroke level, that is, each stroke is processed independently, as shown in \Cref{swsample}. Notably, unlike the S-Down and S-Up, the sampling operations in the DGraph do not affect the structure or features of the SGraph.

\subsection{Information Fusion}

\begin{figure*}[htbp]
\centering
\includegraphics[width=0.70\linewidth]{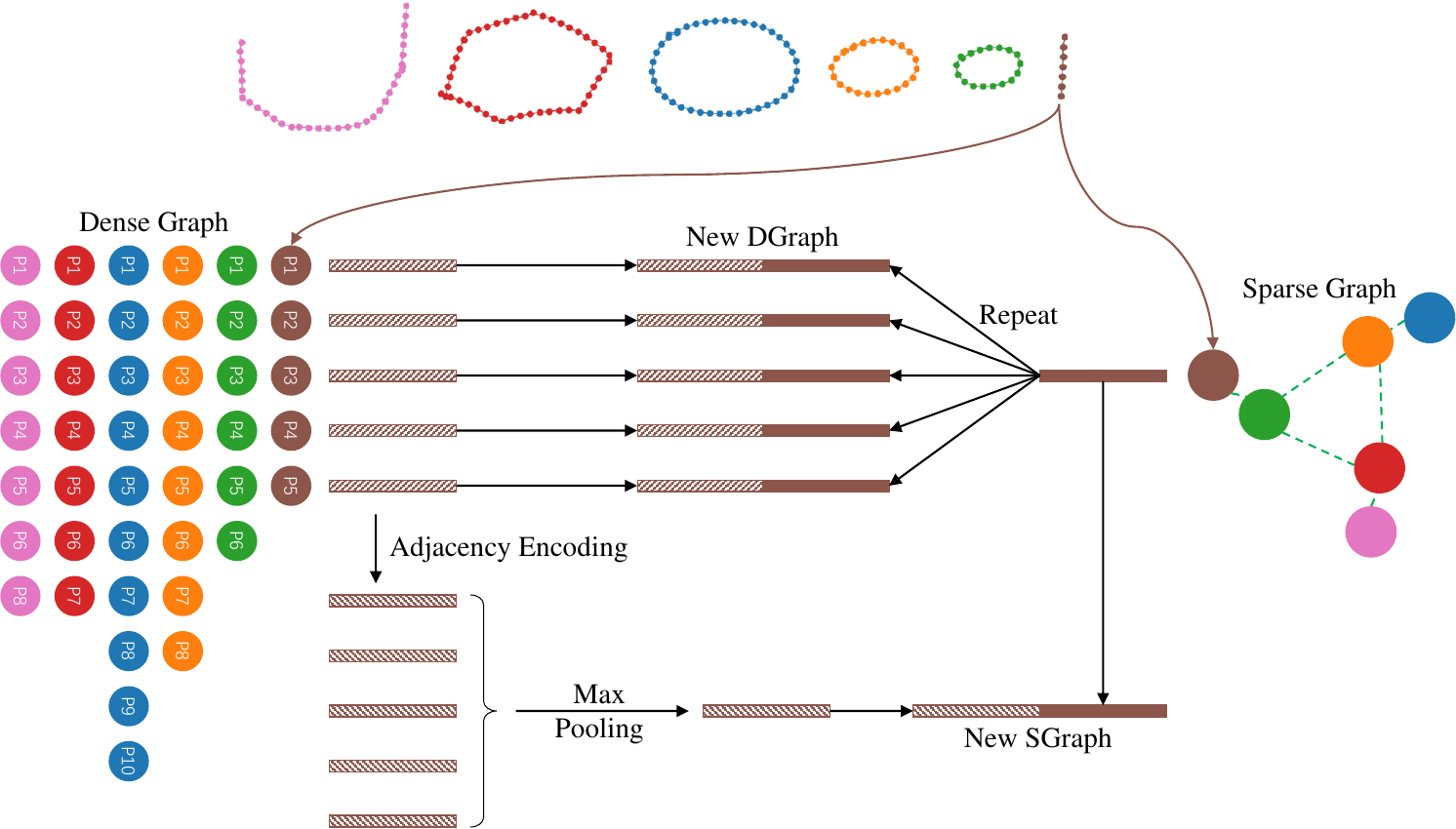}
\caption{\textbf{Information Fusion.} \textit{SGraph features transfer to DGraph}: The SGraph node features are repeated and concatenated to their corresponding DGraph node features (New DGraph). \textit{DGraph features transfer to SGraph}: Adjacency encoding is first performed on the relevant DGraph node features, followed by max-pooling. The resulting features are then concatenated to the corresponding SGraph node features (New SGraph).}
\label{fig:inffusion}
\end{figure*}

The information fusion module facilitates feature transfer and integration between the SGraph and DGraph, as illustrated in \Cref{fig:inffusion}. To transfer features from the SGraph to the DGraph, the SGraph node features are repeated and concatenated to their corresponding DGraph node features. For the DGraph features transfer to the SGraph, the corresponding DGraph nodes for each SGraph node are first identified. These corresponding DGraph nodes are encoded based on the point drawing adjacency; the adjacency encoding module is shown in \Cref{fig:ddownup} (a). The updated SGraph node features are then obtained by max-pooling and feature concatenation.

\subsection{Application Framework}
\begin{figure*}[htbp]
\centering
\includegraphics[width=0.75\linewidth]{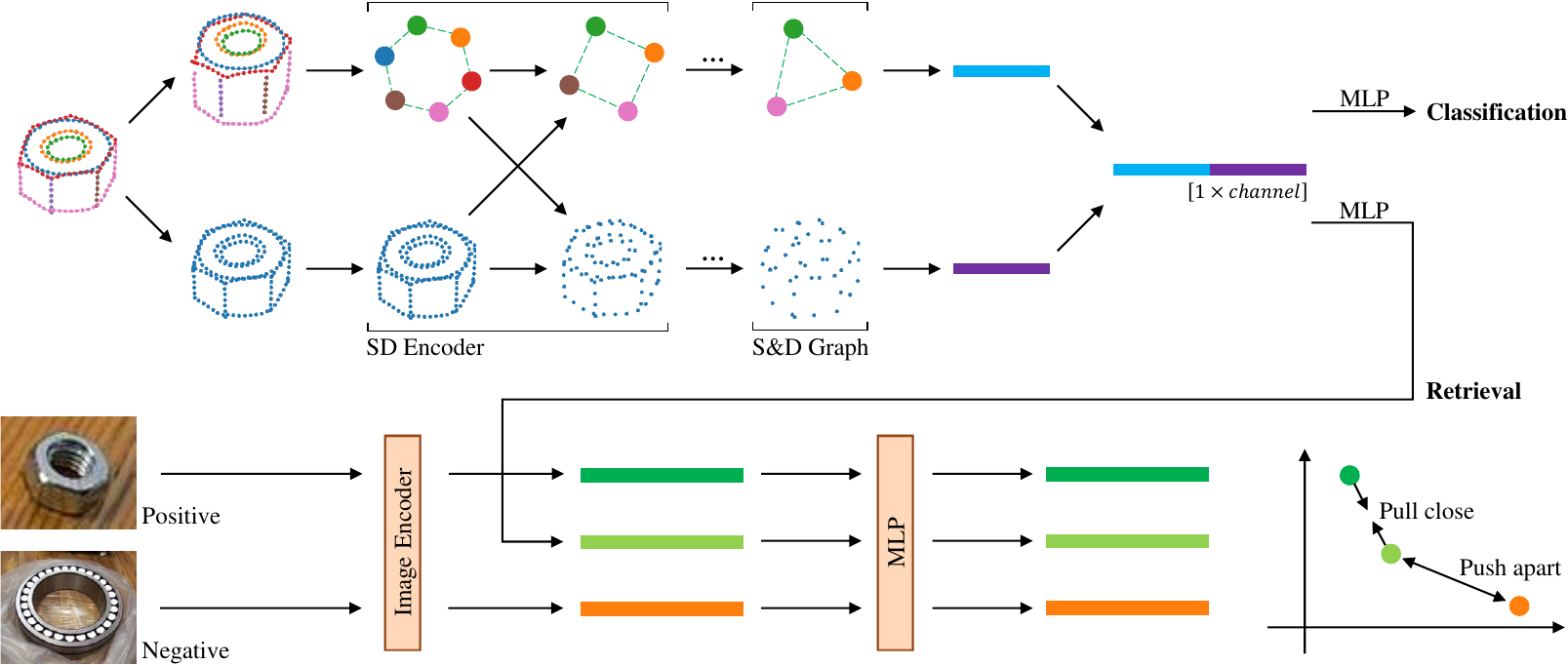}
\caption{\textbf{Vector free-hand sketch classification and sketch based image retrieval framework.} The input vector sketches are first processed through a sequence of SDEncoders. The output S\&D Graphs from the final SDEncoder is aggregated via max-pooling and concatenation to obtain the global feature $\boldsymbol{f}$. For classification, $\boldsymbol{f}$ is fed into MLPs to predict class probabilities. For sketch-based image retrieval, positive and negative image samples are selected carefully: positive samples correspond to the intended retrieval targets, whereas negative samples are visually similar but incorrect. Image features are extracted using an image encoder. The training objective is to minimize the distance between $\boldsymbol{f}$ and its corresponding positive image features, while maximizing the separation from negative image features.}
\label{fig:clsrevl}
\end{figure*}

\textbf{Classification and Retrieval:}
The structure of the classification model is illustrated at the top of \Cref{fig:clsrevl}. After processing by the SD Encoders, the node features from S\&D Graphs are passed through max-pooling to extract global features. These global features are then concatenated and fed into MLPs for classification.
The retrieval model structure is shown at the bottom of \Cref{fig:clsrevl}. The process of extracting global features is identical to that of classification. Global features are then passed through MLPs, and aligned with image features obtained from the pre-trained ULIP \cite{ulip} image encoder. The ULIP encoder is used with fixed pre-trained weights, and a fine-tuning MLP head is added for alignment.

\begin{figure*}[htbp]
\centering
\includegraphics[width=0.85\linewidth]{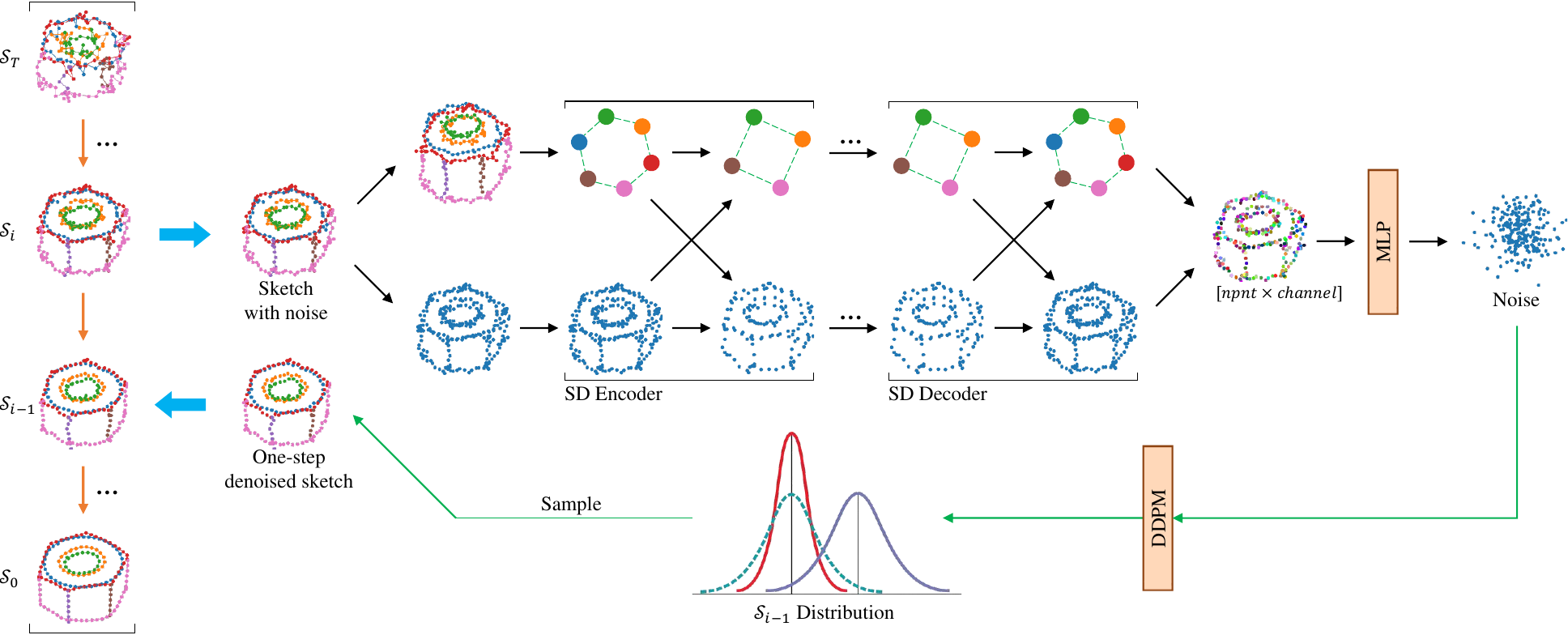}
\caption{\textbf{Vector free-hand sketch generation framework.} This framework is built upon the Denoising Diffusion Probabilistic Model (DDPM) \cite{ddpm}, where SDGraph serves as the noise prediction network. The process starts from pure Gaussian noise $\mathcal{S}_T$ with a predefined maximum number of diffusion steps $T$ (left side). The denoising procedure is then applied iteratively in the reverse direction, from $T$ to $1$. At the $t$-th timestep, the noised sketch $\mathcal{S}_{t}$ and timestep $t$ are fed into SDGraph to predict the noise component $\hat{\boldsymbol{\epsilon}}$. The discrete timestep $t$ is encoded using a sinusoidal positional embedding \cite{trans, ddpm}. Subsequently, the predicted noise $\hat{\boldsymbol{\epsilon}}$, the current noised sketch $\mathcal{S}_{t}$, and the timestep $t$ are passed to the DDPM to estimate the previous state $\mathcal{S}_{t-1}$. After $T$ iterations, the initial Gaussian noise is progressively transformed into a vector free-hand sketch.}
\label{fig:diff}
\end{figure*}

\textbf{Generation:}
The vector freehand sketch generation framework is based on Denoising Diffusion Probabilistic Models (DDPM) \cite{ddpm}, and the timestep is encoded by the SinusoidalPosEmb \cite{sinu}, as shown in \Cref{fig:diff}, where SDGraph serves as the noise prediction module. For input sketches corrupted with noise, both SGraph and DGraph undergo down-sampling and up-sampling operations. Afterward, SGraph features are transferred to DGraph to obtain point-wise features, which are subsequently fed into MLPs to predict the noise.

\section{Experiments}

We evaluate SDGraph across three tasks: sketch classification, sketch-based image retrieval, and sketch generation. All models are optimized using the AdamW optimizer. Experiments are conducted on NVIDIA RTX 4090 GPUs.

\subsection{Classification}
For the classification task, we adopt the QuickDraw dataset \cite{skhrnn}. Due to its large scale, we follow the sampling protocol used in MGT \cite{mgt}, randomly selecting 1,000 samples per class from the 345 categories for training and 100 samples per class for testing. We use the NLL Loss to train SDGraph, as illustrated in \Cref{eq:nll}.

\begin{equation}
\mathcal{L}_{\text{nll}} = - \sum_{i=1}^{C} y_i \log p_i.
\label{eq:nll}
\end{equation}

Where $p_i$ denotes the predicted probability of class $i$ after the Softmax function, $y_i$ is the one-hot encoded ground-truth label, and $C$ is the number of categories.

We compared CNN-, RNN-, and GNN-based methods, with the results summarized in \Cref{cls_res}. CNN-based approaches generally achieve higher accuracy than their RNN- and GNN-based counterparts, indicating the strong capability of CNNs in extracting discriminative features from raster sketches. This performance gap underscores the need for further advancement in learning methods tailored for vector sketch. It is worth noting that vector sketch rasterization inherently eliminates point frequency information and inter-stroke temporal information, which are two factors not considered in this study. In contrast, RNN- and GNN-based methods often utilize these types of information, which may partially account for their relatively lower performance compared to CNN-based methods.

\begin{table}[htbp]
\caption{Classification results on QuickDraw subset\cite{mgt}.}
\centering
\begin{tabularx}{1.0\linewidth}{
    >{\raggedright\arraybackslash}m{0.24\linewidth}  
    >{\raggedright\arraybackslash}m{0.32\linewidth}  
    >{\raggedright\arraybackslash}m{0.12\linewidth}  
    >{\raggedright\arraybackslash}m{0.2\linewidth}  
}

\toprule
\textbf{Architecture \& Network} & & \textbf{Input} & \textbf{Accuracy} \\

\midrule
\multirow[t]{6}{=}{Convolutional Neural NetWorks (CNNs)} & AlexNet \cite{alexnet} & \multirow[t]{6}{=}{Raster sketch} & 0.6808 \\
~ & VGG-19 \cite{vgg} & & 0.6908 \\
~ & Inception V3 \cite{incepnet} & ~ & 0.7422 \\
~ & ResNet-152 \cite{resnet} & ~ & 0.6924 \\
~ & DenseNet-201 \cite{densenet} & ~ & 0.7050 \\
~ & MobileNet V2 \cite{mobil2} & ~ & 0.7310 \\

\midrule
\multirow[t]{4}{=}{Recurrent Neural Networks (RNNs)} & LSTM \cite{lstm} & \multirow[t]{4}{=}{Vector sketch} & 0.6068 \\
~ & SketchRNN \cite{skhrnn} & ~ & 0.6665 \\
~ & GRU \cite{gru} & ~ & 0.6224 \\
~ & Bi-Dir. GRU \cite{gru} & ~ & 0.6768 \\

\midrule
\multirow[t]{7}{=}{Graph Neural Networks (GNNs)} & GCN \cite{gcn} & \multirow[t]{7}{=}{Vector sketch} & 0.6800 \\
~ & GAT \cite{gat} & ~ & 0.6977 \\
~ & Vanilla Transformer \cite{trans} & ~ & 0.5249 \\
~ & MGT (Base) \cite{mgt} & ~ & 0.7070 \\
~ & MGT (Large) \cite{mgt} & ~ & 0.7280 \\
~ & SketchTransformer \cite{skhtrans} & ~ & 0.6829 \\
~ & \textbf{Ours} & ~ & \textbf{0.7537} \\

\bottomrule
\end{tabularx}
\label{cls_res}
\end{table}

\subsection{Retrieval}

We evaluated SDGraph on three retrieval tasks: category-level zero-shot sketch-based image retrieval (CL-ZS-SBIR), fine-grained zero-shot sketch-based image retrieval (FG-ZS-SBIR), and fine-grained sketch-based image retrieval (FG-SBIR). The differences between the above three retrieval tasks are shown in \Cref{tab:retrieval_task}.

\begin{table}[t]
\centering
\caption{Comparison of different SBIR task settings. $\mathcal{C}^{T}$: category of training set, $\mathcal{C}^{E}$: category of testing set.}
\begin{tabularx}{1.0\linewidth}{
    >{\raggedright\arraybackslash}m{0.275\linewidth}  
    >{\raggedright\arraybackslash}m{0.365\linewidth}  
    >{\raggedright\arraybackslash}m{0.35\linewidth}  
}
\toprule
Task & Target & Dataset relation \\
\midrule
CL-ZS-SBIR & class-level match & $\mathcal{C}^{T} \cap \mathcal{C}^{E} = \emptyset$ \\
FG-ZS-SBIR & instance-level match & $\mathcal{C}^{T} \cap \mathcal{C}^{E} = \emptyset$ \\
FG-SBIR & instance-level match & $\mathcal{C}^{T} = \mathcal{C}^{E}$ \\
\bottomrule
\end{tabularx}
\label{tab:retrieval_task}
\end{table}

The task of CL-ZS-SBIR is defined as: given a sketch $\mathcal{S}_i$ belonging to class $i$, retrieve pictures ${\mathcal{P}}_{i,j}$ belonging to class $i$, from picture set $\left\{\mathcal{P}_{i,j}\right\}{_{i=1}^{N_c}|_{j=1}^{N_i}}$ which contains $N_c$ classes, and class $i$ containing $N_i$ pictures. The training classes $\mathcal{C}^T=\left\{c_1^T,c_2^T,\cdots,c_N^T\right\}$ and the testing classes $\mathcal{C}^E=\left\{c_1^E,c_2^E,\cdots,c_M^E\right\}$ have no intersection, i.e., $\mathcal{C}^T \cap \mathcal{C}^E = \emptyset$. Therefore, the goal of CL-ZS-SBIR is to learn features that bring sketches and images of the same class closer together in the embedding space, while pushing apart those of different classes. In this context, negative samples are images that do not belong to the same class as the query sketch.

We use the Sketchy-extend dataset \cite{sketchy} to evaluate the performance of CL-ZS-SBIR. The Sketchy-extend dataset \cite{sketchy} contains 125 categories, of which 104 are used for training, and the remaining 21 are reserved for testing in the zero-shot setting, following the class split protocol established by ZSE-SBIR \cite{zse}. 
During training, positive samples are defined as images belonging to the same category as the query sketch, while negative samples are drawn from different categories. The SDGraph is optimized using the Triplet Loss \cite{triplet_loss}, as formulated in \Cref{eq:triplet_loss}.


\begin{equation}
\begin{aligned}
\mathcal{L}_{\text{tri}}
&=
-\frac{1}{|\mathcal{P}(i)|}
\sum_{j \in \mathcal{P}(i)}
\log
\frac{
\exp \big( \mathrm{sim}(\boldsymbol{s}_i, \boldsymbol{p}_{i,j}) / \tau \big)
}{
\sum_{\substack{\boldsymbol{a} \in \mathcal{A}(i)}}
\exp \big( \mathrm{sim}(\boldsymbol{s}_i, \boldsymbol{a}) / \tau \big)
}
\label{eq:triplet_loss}
\end{aligned}
\end{equation}

Where $\boldsymbol{s}_i$ denotes the $i$-th query sketch feature. 
$\mathcal{P}(i)$ represents the positive image set associated with $\boldsymbol{s}_i$, and $|\mathcal{P}(i)|$ denotes the number of samples in $\mathcal{P}(i)$.
$\mathcal{A}(i)$ is the contrastive set for $\boldsymbol{s}_i$, which consists of all positive samples in $\mathcal{P}(i)$ together with all negative samples.
$\boldsymbol{p}_{i,j}$ represents the $j$-th positive image feature associated with $\boldsymbol{s}_i$, 
The function $\mathrm{sim}(\cdot,\cdot)$ computes the cosine similarity between
two features, defined as
$\mathrm{sim}(\boldsymbol{x}, \boldsymbol{y}) = \boldsymbol{x}^{\top}\boldsymbol{y}/\left(\lVert\boldsymbol{x} \rVert \cdot \lVert \boldsymbol{y} \rVert\right)$.
The temperature parameter $\tau$ controls the sharpness of the similarity
distribution.

The results of CL-ZS-SBIR are presented in \Cref{tab:cl_sbir}, where the CNN-, RNN-, and GNN-based methods are compared. In general, methods that take vector sketches as input tend to achieve lower accuracy compared with those using raster sketches, which may result from the closer modality between rasterized sketches and images. Despite this, our SDGraph achieves the best overall performance. Although the improvement over CNN-based methods is marginal, SDGraph demonstrates a significant performance improvement compared with other methods which based on vector sketch input. 

\begin{figure*}[htbp]
\centering
\includegraphics[width=1.0\linewidth]{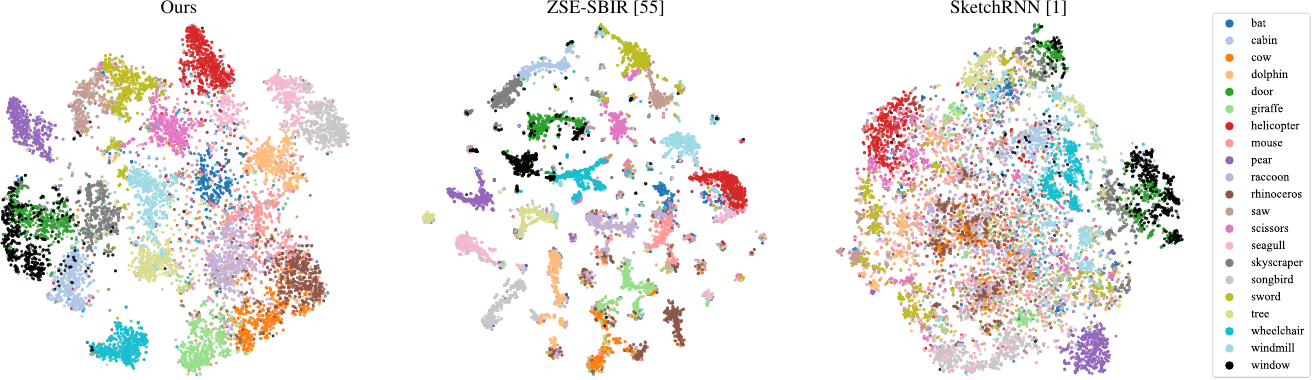}
\caption{\textbf{Feature visualization of CL-ZS-SBIR.} The features are extracted from testing categories $\mathcal{C}^E=\left\{c_1^E,c_2^E,\cdots,c_{21}^E\right\}$, where $\mathcal{C}^E$ has no intersection with the training categories $\mathcal{C}^T=\left\{c_1^T,c_2^T,\cdots,c_{104}^T\right\}$, i.e., $\mathcal{C}^T \cap \mathcal{C}^E = \emptyset$.}
\label{fig:cl_vis}
\end{figure*}

The feature visualizations are shown in \Cref{fig:cl_vis}. Since the test categories are unseen during training, features from the test sketch set are based on knowledge learned from training categories, leading to some class-level dispersion. While ZSE-SBIR \cite{zse} exhibits distinct separation between classes, there is significant overlap among categories and a large number of clusters, indicating weak intra-class compactness. In contrast, although the cluster separation of SDGraph is less pronounced, the number of clusters is lower than that of ZSE-SBIR, demonstrating SDGraph’s stronger cross-category generalization ability.

\begin{table}[htbp]
\caption{Results of CL-ZS-SBIR on Sketchy-extend dataset \cite{sketchy}.}
\centering
\begin{tabularx}{1.0\linewidth}{
    >{\raggedright\arraybackslash}m{0.24\linewidth}  
    >{\raggedright\arraybackslash}m{0.24\linewidth}  
    >{\raggedright\arraybackslash}m{0.11\linewidth}  
    >{\raggedright\arraybackslash}m{0.1\linewidth}  
    >{\raggedright\arraybackslash}m{0.1\linewidth}  
}

\toprule
\textbf{Architecture \& Network} & & \textbf{Input} & \textbf{\shortstack{mAP\\@200}} & \textbf{\shortstack{Prec.\\@200}} \\

\midrule
\multirow[t]{8}{=}{Convolutional Neural NetWorks (CNNs)} & SAKE \cite{sake} & \multirow[t]{8}{=}{Raster sketch} & 0.497 & 0.598 \\
~ & IIAE \cite{iiae} & & 0.373 & 0.485 \\
~ & CAAE \cite{cavae} & & 0.156 & 0.260 \\
~ & CVAE \cite{cavae} & & 0.225 & 0.333 \\
~ & GRL \cite{grl} & & 0.369 & 0.370 \\
~ & LVM \cite{lvm} & & 0.723 & 0.725 \\
~ & ZSE-SBIR \cite{zse} & & 0.520 & 0.617 \\
~ & Sketch3T \cite{sketch3t} & & 0.579 & 0.648 \\
~ & SD-PL \cite{sdpl} & & 0.746 & 0.747 \\

\midrule
\multirow[t]{4}{=}{Recurrent Neural Networks (RNNs)} & LSTM \cite{lstm} & \multirow[t]{4}{=}{Vector sketch} & 0.393 & 0.412 \\
~ & SketchRNN \cite{skhrnn} & ~ & 0.471 & 0.489 \\
~ & GRU \cite{gru} & ~ & 0.409 & 0.433 \\
~ & Bi-Dir. GRU \cite{gru} & ~ & 0.455 & 0.471 \\

\midrule
\multirow[t]{3}{=}{Graph Neural Networks (GNNs)} & GCN \cite{gcn} & \multirow[t]{3}{=}{Vector sketch} & 0.493 & 0.510 \\
~ & GAT \cite{gat} & ~ & 0.508 & 0.525 \\
~ & \textbf{Ours} & ~ & \textbf{0.763} & \textbf{0.770} \\

\bottomrule
\end{tabularx}
\label{tab:cl_sbir}
\end{table}

The task of FG-ZS-SBIR aims to achieve instance-level matching between sketches and images. Similar to CL-ZS-SBIR, the training classes $\mathcal{C}^T$ and the testing classes $\mathcal{C}^E$ are disjoint, i.e., $\mathcal{C}^T \cap \mathcal{C}^E = \emptyset$, ensuring zero-shot generalization across unseen categories. Therefore, the objective of FG-ZS-SBIR is to learn features that bring instance-level matched sketch-image pairs as close as possible in the embedding space. In this setting, negative samples refer to images that belong to the same category as the query sketch, but do not match at the instance level.


We use the Sketchy-extend dataset \cite{sketchy} to evaluate the performance of FG-ZS-SBIR. The SDGraph is optimized by the Triplet Loss in \Cref{eq:triplet_loss} and the Relation Loss \cite{zse}, as formulated in \Cref{eq:re_loss}.

\begin{equation}
\begin{aligned}
\mathcal{L} &= \mathcal{L}_\text{tri} + \lambda \cdot \mathcal{L}_\text{re}\\
\mathcal{L}_\text{re} &= \mathrm{MSE}\left(\boldsymbol{s}_i, \boldsymbol{p}_i \right) \\
\label{eq:re_loss}
\end{aligned}
\end{equation}


Where $\boldsymbol{p}_i$ denotes the instance-level matched image feature corresponding to the query sketch. In the FG-ZS-SBIR settings, each query sketch is associated with only one positive sample in the Triplet Loss \Cref{eq:triplet_loss}, i.e., $|\mathcal{P}(i)| = 1$.

\begin{figure*}[htbp]
\centering
\includegraphics[width=1.0\linewidth]{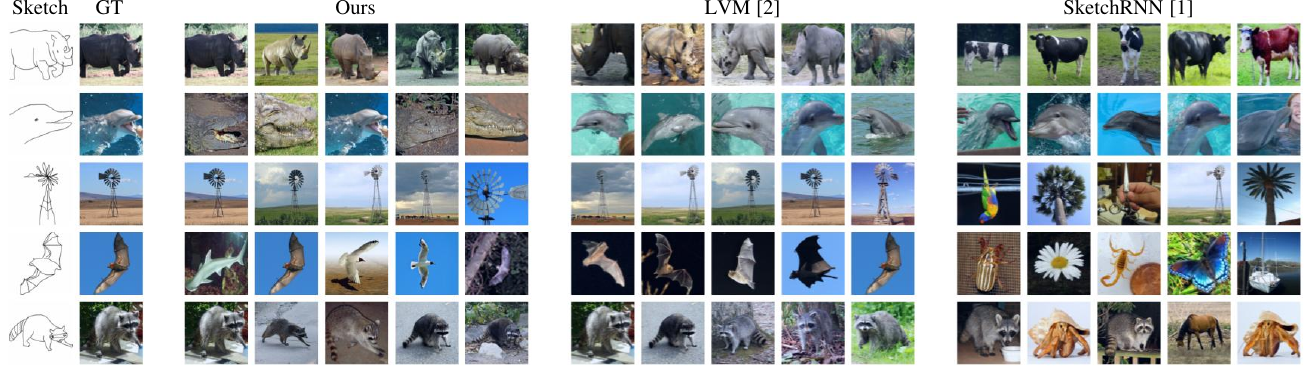}
\caption{\textbf{Representative FG-ZS-SBIR results across GNN-, CNN-, RNN-based methods.} The Ground Truth (GT) refers to the instance-level paired image corresponding to the query sketch.}
\label{fg_vis}
\end{figure*}

The results of FG-ZS-SBIR are reported in \Cref{tab:fg_zs_sbir}, from which we observe conclusions consistent with those of CL-ZS-SBIR. Fine-grained retrieval examples are shown in \Cref{fg_vis}. It can be seen that SDGraph pays more attention to the overall contour structure of the retrieved images. For instance, the retrieved images of rhinoceroses all face to the right, dolphins are mistakenly retrieved as crocodiles due to similar outlines, and raccoons are consistently retrieved with paws extended to the right. In contrast, LVM focuses more on the semantic content of the sketch. For example, although all retrieved images for the rhino sketch belong to the rhino category, most of them face left, and their global outlines do not align well with the query sketch.

\begin{table}[htbp]
\caption{Results of FG-ZS-SBIR on Sketchy-extend dataset \cite{sketchy}.}
\centering
\begin{tabularx}{1.0\linewidth}{
    >{\raggedright\arraybackslash}m{0.24\linewidth}  
    >{\raggedright\arraybackslash}m{0.23\linewidth}  
    >{\raggedright\arraybackslash}m{0.09\linewidth}  
    >{\raggedright\arraybackslash}m{0.1\linewidth}  
    >{\raggedright\arraybackslash}m{0.1\linewidth}  
}

\toprule
\textbf{Architecture \& Network} & & \textbf{Input} & \textbf{\shortstack{Acc.@1}} & \textbf{\shortstack{Acc.@5}} \\

\midrule
\multirow[t]{3}{=}{Convolutional Neural NetWorks (CNNs)} & Gen-VAE \cite{genvae} & \multirow[t]{3}{=}{Raster sketch} & 0.226 & 0.490 \\
~ & Grad-VAE \cite{gradvae} & & 0.134 & 0.349 \\
~ & LVM \cite{lvm} & & 0.287 & 0.623 \\
~ & SD-PL \cite{sdpl} & & 0.319 & 0.658 \\

\midrule
\multirow[t]{4}{=}{Recurrent Neural Networks (RNNs)} & LSTM \cite{lstm} & \multirow[t]{4}{=}{Vector sketch} & 0.135 & 0.279 \\
~ & SketchRNN \cite{skhrnn} & ~ & 0.156 & 0.339 \\
~ & GRU \cite{gru} & ~ & 0.141 & 0.293 \\
~ & Bi-Dir. GRU \cite{gru} & ~ & 0.162 & 0.344 \\

\midrule
\multirow[t]{3}{=}{Graph Neural Networks (GNNs)} & GCN \cite{gcn} & \multirow[t]{3}{=}{Vector sketch} & 0.174 & 0.403 \\
~ & GAT \cite{gat} & ~ & 0.180 & 0.441 \\
~ & \textbf{Ours} & ~ & \textbf{0.328} & \textbf{0.669} \\

\bottomrule
\end{tabularx}
\label{tab:fg_zs_sbir}
\end{table}

The task of FG-SBIR is similar to FG-ZS-SBIR; the difference lies in the training categories $\mathcal{C}^T$ are the same as the testing categories $\mathcal{C}^E$ for FG-SBIR, i.e., $\mathcal{C}^T = \mathcal{C}^E$.
We use the QMUL\_ShoeV2/ChairV2 dataset \cite{shoe_chair} to evaluate FG-SBIR; the loss function is the same as FG-ZS-SBIR.


The results of FG-SBIR are listed in \Cref{tab:fg_sbir}, compared with FG-ZS-SBIR in \Cref{tab:fg_zs_sbir}, FG-SBIR achieves significantly higher performance in terms of Acc@K. Beyond the inherent differences in the datasets, the primary reason is that the FG-SBIR shares the same category space between the training and testing, resulting in a smaller distribution gap. In contrast, FG-ZS-SBIR adopts a zero-shot setting where the training and testing categories are completely disjoint, leading to a substantially larger distribution shift, which significantly increases the retrieval difficulty and consequently leads to lower Acc@K. 

\begin{table}[htbp]
\caption{Results of FG-SBIR on QMUL\_Shoe/Chair dataset \cite{shoe_chair}. From top to bottom, the methods are based on CNN (raster sketch), RNN (vector sketch), and GNN (vector sketch).}
\centering
\setlength{\tabcolsep}{8pt}
\begin{tabularx}{1.0\linewidth}
{
    >{\raggedright\arraybackslash}m{0.28\linewidth}  
    >{\centering\arraybackslash}m{0.1\linewidth}  
    >{\centering\arraybackslash}m{0.1\linewidth}  
    >{\centering\arraybackslash}m{0.1\linewidth}  
    >{\centering\arraybackslash}m{0.1\linewidth}  
}
\toprule
\raisebox{-0.7ex}{\multirow{2}{*}{\textbf{Methods}}} &
\multicolumn{2}{c}{\textbf{ChairV2}} & \multicolumn{2}{c}{\textbf{ShoeV2}} \\
\cmidrule(lr){2-3} \cmidrule(lr){4-5}
& Acc.@1 & Acc.@5 & Acc.@1 & Acc.@5 \\
\midrule
Triplet-SN \cite{triplet_sn}  & 0.474 & 0.714 & 0.287 & 0.635 \\
HOLEF-SN \cite{holef_sn}      & 0.507 & 0.736 & 0.312 & 0.666 \\
Partial-OT \cite{partial_ot}  & 0.633 & 0.797 & 0.399 & 0.682 \\
CrossHier \cite{crosshier}    & 0.624 & 0.791 & 0.362 & 0.678 \\
StyleMeUp \cite{style_me_up}  & 0.628 & 0.793 & 0.304 & 0.610 \\
On-the-fly \cite{on_the_fly}  & 0.512 & 0.738 & 0.368 & 0.685 \\
SketchPVT \cite{sketchpvt}    & 0.712 & 0.801 & 0.441 & 0.708 \\
\midrule
LSTM \cite{lstm} & 0.220 & 0.365 & 0.153 & 0.299 \\  
SketchRNN \cite{skhrnn} & 0.232 & 0.375 & 0.158 & 0.306 \\
GRU \cite{gru} & 0.235 & 0.384 & 0.164 & 0.315 \\
Bi-Dir. GRU \cite{gru} & 0.238 & 0.406 & 0.176 & 0.351 \\
\midrule
GCN \cite{gcn} & 0.254 & 0.437 & 0.179 & 0.354 \\
GAT \cite{gat} & 0.251 & 0.440 & 0.190 & 0.389 \\
\textbf{Ours} & \textbf{0.724} & \textbf{0.805} & \textbf{0.455} & \textbf{0.715} \\
\bottomrule
\end{tabularx}
\label{tab:fg_sbir}
\end{table}

\begin{figure*}[htbp]
\centering
\includegraphics[width=1.0\linewidth]{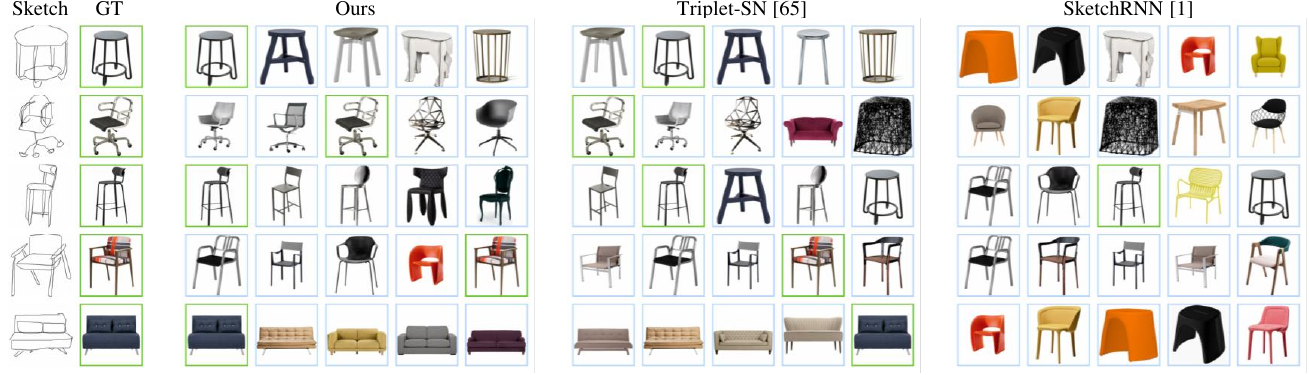}
\caption{\textbf{Representative FG-SBIR results on QMUL-ChairV2 dataset.} The QMUL-ChairV2 \cite{shoe_chair} dataset contains 1275 sketches and 400 photos, from which 952 sketches and 300 photos are used for training, and the rest 323 sketches and 100 photos are used for testing.}
\label{fg_vis_chair}
\end{figure*}

\begin{figure*}[htbp]
\centering
\includegraphics[width=1.0\linewidth]{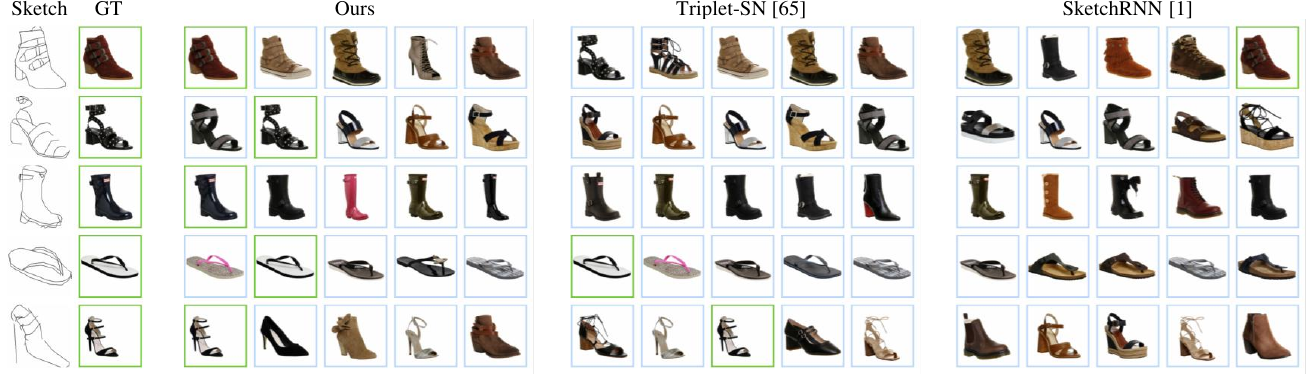}
\caption{\textbf{Representative FG-SBIR results on QMUL-ShoeV2 dataset.} The QMUL-ShoeV2 \cite{shoe_chair} dataset contains 6648 sketches and 2000 photos, from which 5982 sketches and 1800 photos are used for training, the rest 666 sketches and 200 photos are used for testing.}
\label{fig:fg_vis_shoe}
\end{figure*}

The retrieval examples for FG-SBIR are illustrated in \Cref{fg_vis_chair} and \Cref{fig:fg_vis_shoe}. Compared with the FG-ZS-SBIR results shown in \Cref{fg_vis}, the most notable difference is that FG-SBIR rarely retrieves category-level mismatched images. For example, a sketch of a high heel is unlikely to retrieve knee-high boots (\Cref{fig:fg_vis_shoe}, fifth row). In contrast, FG-ZS-SBIR often retrieves images that are structurally aligned with the query sketch but belong to different categories. A representative example is that a dolphin sketch is mistakenly matched to a crocodile with its mouth open (\Cref{fg_vis}, second row). This behavior arises because the testing categories in FG-ZS-SBIR are unseen during training, leaving the model without access to category-specific texture or appearance cues. In FG-SBIR, where the training and testing sets share the same category space, such mismatches are largely avoided.

\subsection{Generation}

The QuickDraw dataset \cite{skhrnn} is adopted for generation, in which six categories are selected and categorized into simple, medium, and complex groups according to their typical numbers of strokes and structural details. During training, 70,000 samples per category were selected from the “train” subset. For evaluation, the Fréchet Inception Distance (FID) was computed using 1,000 samples per category from the “test” subset. The FID scores were calculated using the pytorch\_fid \cite{fid} implementation with dims=2048, and all color images were converted into binary black-and-white images before computing the FID. 

We train SDGraph following the DDPM paradigm. 
Given a clean sketch $\mathcal{S}_0$, a forward diffusion process is defined 
by progressively corrupting the Gaussian noise over $T$ timesteps. 
At an arbitrary timestep $t \in \{1,\dots,T\}$, the noised sketch $\mathcal{S}_t$ is obtained as:


\begin{equation}
\begin{aligned}
\mathcal{S}_t &= \sqrt{\bar{\alpha}_t}\cdot\mathcal{S}_0 + \sqrt{1-\bar{\alpha}_t}\cdot\boldsymbol{\epsilon}, 
\quad \boldsymbol{\epsilon} \sim \mathcal{N}(\mathbf{0},\mathbf{I}) 
\label{eq:add_noise}
\end{aligned}
\end{equation}

Where $\bar{\alpha}_t=\prod_{i=1}^{t}(1-\beta_i)$, and $\{\beta_t\}$ is a predefined noise schedule. $\beta_t = \beta_{\min} + \frac{t-1}{T-1}(\beta_{\max} - \beta_{\min}), \beta_{\min}=0, \beta_{\max}=0.02, T=1000$ in this paper.

During training, $\mathcal{S}_t$ is first obtained by \Cref{eq:add_noise}, where both $\mathcal{S}_0$ and timestep $t$ are randomly selected. $\mathcal{S}_t$ and $t$ are then fed into SDGraph to predict the injected noise:


\begin{equation}
\hat{\boldsymbol{\epsilon}} = \boldsymbol{\epsilon}_\theta(\mathcal{S}_t, t),
\end{equation}

Where $\boldsymbol{\epsilon}_\theta(\cdot)$ denotes the parameter of SDGraph.

The training objective is to minimize the mean squared error between the predicted noise $\hat{\boldsymbol{\epsilon}}$
and the ground-truth noise $\boldsymbol{\epsilon}$:
\begin{equation}
\mathcal{L}_{\text{diff}} = 
\mathbb{E}_{\mathcal{S}_0,\, t,\, \boldsymbol{\epsilon}}
\left[
\left\| \boldsymbol{\epsilon} - \hat{\boldsymbol{\epsilon}} \right\|_2^2
\right].
\end{equation}

\begin{figure*}[htbp]
\centering
\includegraphics[width=1.0\linewidth]{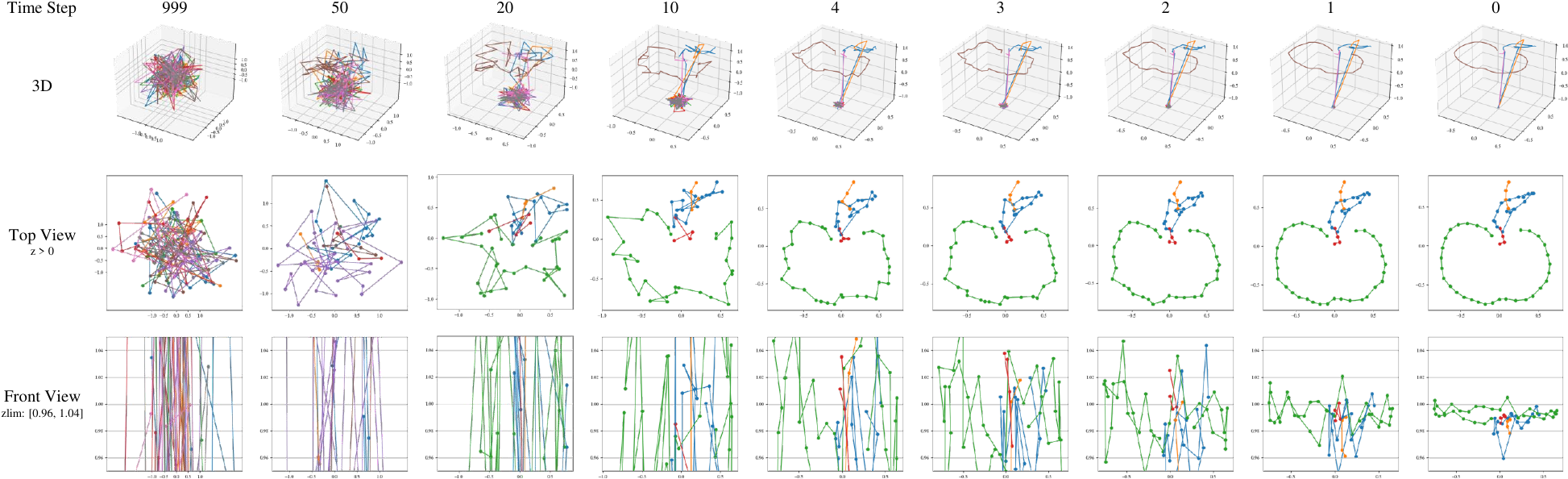}
\caption{\textbf{Sketch generation process.} Starting from pure noise at timestep 999, the sketch gradually evolves to timestep 0. Points moving toward $z=1$ form valid sketch points, while those moving toward $z=-1$ serve as padding. Points with $z<0$ are treated as invalid and removed during post-processing, enabling sketches with varying numbers of strokes and points.}
\label{fig:gen_prog}
\end{figure*}

The generation process is shown in \Cref{fig:gen_prog}, which starts from a tensor of size $[m, n, 3]$, where each element is independently sampled from a Gaussian distribution $\mathcal{N}(0, 1)$. Here, $m$ denotes the predefined maximum number of strokes and $n$ denotes the maximum number of points per stroke.
During denoising, point coordinates are progressively updated, and the adjacency matrix of SDGraph is dynamically constructed via KNN. As the point positions evolve across time steps, the graph structure correspondingly changes.
In our representation, padded points are defined as $(0,0,-1)$, while valid sketch points are represented as $(x,y,1)$. During the denoising process, some points gradually move toward the $z=1$ plane, while others move toward the $z=-1$ plane, implicitly determining their validity.
Consequently, the number of valid points varies dynamically during generation, allowing both the number of strokes and the number of points within each stroke to change adaptively, thereby improving the diversity of generated sketches.

We compared SDgraph with several representative vector sketch generation approaches covering different methodological paradigm. The FID scores of the compared methods are reported in \Cref{tab:fid}. From other methods, although SketchKnitter \cite{skhknitter} also adopts the diffusion pipeline, its FID scores are consistently higher than those of SDGraph. While SP-gra2seq \cite{spgra} achieves comparable FID values to SDGraph on Apple and Bicycle categories, it shows substantially larger performance gaps on Shark and Angel categories. Overall, these results demonstrate that SDGraph possesses superior modeling capability across different levels of sketch complexity.

\newcommand\myvalue{0.068}
\begin{table}[htbp]
\caption{FID$\downarrow$ of the generated sketches on QuickDraw \cite{skhrnn}.}
\centering
\begin{tabularx}{1.0\linewidth}{
    >{\raggedright\arraybackslash}m{0.24\linewidth} 
    >{\centering\arraybackslash}m{\myvalue\linewidth} 
    >{\centering\arraybackslash}m{\myvalue\linewidth} 
    >{\centering\arraybackslash}m{\myvalue\linewidth} 
    >{\centering\arraybackslash}m{\myvalue\linewidth} 
    >{\centering\arraybackslash}m{\myvalue\linewidth} 
    >{\centering\arraybackslash}m{\myvalue\linewidth} 
}

\toprule
\raisebox{-0.7ex}{\multirow{2}{=}{\textbf{Method}}} & \multicolumn{2}{c}{\textbf{Simple}} & \multicolumn{2}{c}{\textbf{Medium}} & \multicolumn{2}{c}{\textbf{Complex}} \\

\cmidrule{2-7}
~ & apple & moon & book & shark & angel & bicycle \\
\midrule

SketchKnitter \cite{skhknitter} & 271.4 & 278.2 & 263.4 & 268.9 & 302.0 & 290.1 \\
SketchHealer \cite{sketchhealer} & 102.5 & 96.5 & 53.4 & 72.8 & 90.2 & 120.2 \\
SketchRNN \cite{skhrnn} & 94.7 & 130.2 & 74.5 & 79.3 & 194.1 & 125.9 \\
SketchLattice \cite{sketchlattice} & 52.7 & 91.8 & 49.7 & 122.4 & 90.2 & 102.7 \\
DC-gra2seq \cite{dcgra} & 39.6 & 72.1 & 48.4 & 50.1 & 88.4 & 72.4 \\
SP-gra2seq \cite{spgra} & 37.3 & 69.7 & 47.6 & 66.9 & 62.6 & 54.2 \\

Ours & \textbf{36.2} & \textbf{60.0} & \textbf{47.3} & \textbf{33.6} & \textbf{48.6} & \textbf{53.3} \\

\bottomrule
\end{tabularx}
\label{tab:fid}
\end{table}

\begin{figure*}[htbp]
\centering
\includegraphics[width=1.0\linewidth]{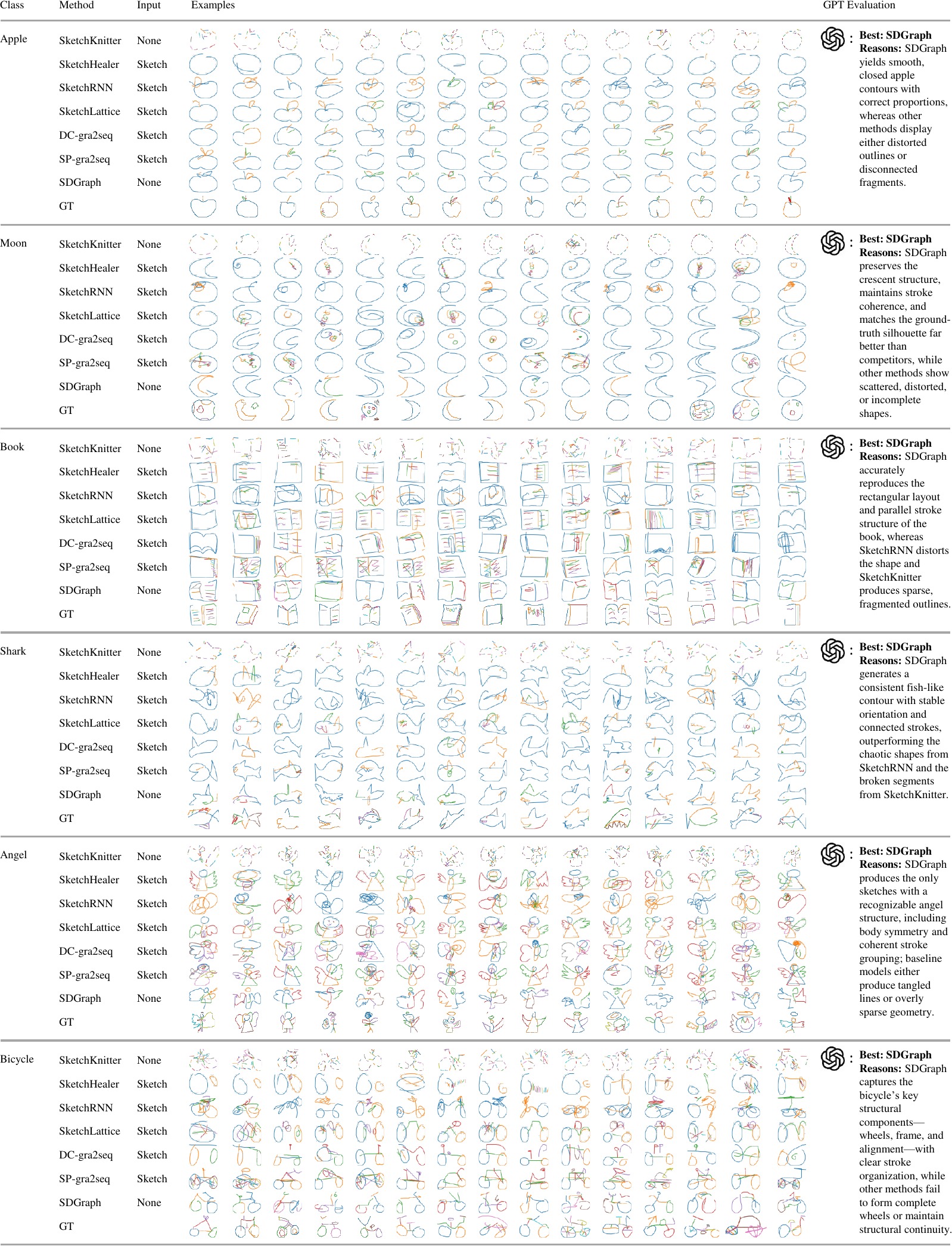}
\caption{\textbf{Generated sketches and ChatGPT evaluation.} We ask ChatGPT “Based on the sketches generated by each method, select the method with the highest generation quality and provide the reasons.” to obtain evaluations. All methods except SDGraph and SketchKnitter use sketches as input to improve generation quality and stability. The input sketches are randomly sampled from the QuickDraw “valid” subset.}
\label{fig:gen_vis}
\end{figure*}

Examples of generated sketches are presented in \Cref{fig:gen_vis}, where the “GT” row shows instances randomly sampled from the QuickDraw “test” subset. In addition, we conducted a qualitative assessment by requesting ChatGPT to evaluate the generated sketches. From results in \Cref{fig:gen_vis}, it can be observed that the sketches generated by SketchKnitter tend to contain a large number of spatially dispersed strokes, suggesting weak control over stroke continuity (SketchKnitter is reproduced by the official code\footnote{\url{https://github.com/wangqiang9/SketchKnitter}}).
The fragmented and broken strokes may arise from the design of SketchKnitter’s pen-state prediction module. Specifically, SketchKnitter predicts the pen state by applying MLPs to the point-wise features produced by the denoising network. However, these features are primarily optimized for noise estimation in the diffusion process, and may not sufficiently encode stroke-level geometric continuity. In contrast, SDGraph explicitly groups points belonging to the same stroke into a single structured representation, thereby avoiding this issue. 

The SketchRNN performs relatively well when generating sketches with a small number of strokes, but struggles to produce coherent results for sketches with a higher number of strokes, often resulting in single-stroke outputs. Additionally, SketchRNN is weak in generating strokes with cusps. For example, in the moon category, it predominantly generates full moon shapes, with only a few instances resembling string moons. 

SketchHealer, SketchLattice, and DC-gra2seq show improved generation quality compared with SketchRNN; nevertheless, they still exhibit noticeable artifacts such as spiral structures and repetitive strokes in certain categories (e.g., Moon and Angel). SP-gra2seq achieves visually appealing results overall, but its point-wise pen-state prediction remains suboptimal. For example, in categories such as Moon and Book, strokes that should be segmented at intermediate points are often generated as continuous strokes, leading to incorrect stroke boundaries.

The rightmost column reports the ChatGPT evaluation results, which indicate that SDGraph generates vector freehand sketches perceived to be of higher quality than those produced by other investigated methods.

\subsection{Ablation Studies}

The ablation studies are mainly conducted on the classification task because it provides a direct and quantitative evaluation of the learned sketch representations. The retrieval model shares a similar architecture with the classification model, and improvements in classification performance are therefore generally reflected in retrieval performance. For sketch generation, there is currently no widely accepted quantitative metric for evaluating structural plausibility, and generation quality is typically assessed through qualitative comparisons.

To validate the effectiveness of our proposed design, we conducted a series of ablation studies focusing on the following modules. 
The experiments were performed on the subset of QuickDraw \cite{skhrnn}, and the results are reported in \Cref{ablation}.

\begin{enumerate}
    \item SGraph (SG).
    \item DGraph (DG).
    \item Stroke Sample in SGraph (SS).
    \item Point Sample in DGraph (PS).
    \item Information Fusion (IF).
\end{enumerate}

\noindent\textbf{Effectiveness of SGraph and DGraph:}

\begin{itemize}
    \item SG vs. DG
\end{itemize}

The comparison between SG and DG demonstrates that DGraph achieves higher accuracy than SGraph, suggesting that the local features extracted by DGraph are more informative than the stroke features captured by SGraph. However, SGraph offers approximately five times faster inference speed, primarily due to its significantly fewer nodes. It is also worth noting that despite the smaller number of nodes, SGraph contains more parameters than DGraph, as each node in SGraph has a longer feature representation.

\noindent\textbf{Effectiveness of combining SGraph and DGraph:}

\begin{itemize}
    \item SG, DG vs. (SG + DG)
    \item SG, (DG + PS) vs. SG + (DG + PS)
    \item (SG + SS), DG vs. (SG + SS) + DG
    \item (SG + SS), (DG + PS) vs. (SG + SS) + (DG + PS)
\end{itemize}

The comparison among SG, DG, and the combined (SG + DG) reveals that integrating SGraph and DGraph improves overall accuracy, highlighting the complementary strengths of stroke-level features from SGraph and point-level features from DGraph. Furthermore, the final three comparisons demonstrate that even after applying sampling operations, the combination of SGraph and DGraph consistently leads to improved accuracy, while introducing only a marginal increase in inference time.

\noindent\textbf{Effectiveness of Node Sample:}

\begin{itemize}
    \item SG vs. (SG + SS)
    \item DG vs. (DG + PS)
    \item (SG + SS) vs. (SG + SS) + DG
    \item (DG + PS) vs.SG + (DG + PS)
    \item SG + DG vs. (SG + SS) + (DG + PS)
\end{itemize}

Comparisons between SG and (SG + SS), as well as DG and (DG + PS), indicate that the sampling modules S-Down and D-Down contribute to accuracy improvements when applied individually to SGraph and DGraph. While the inclusion of these sampling modules increases the number of parameters, their impact on inference time differs. Specifically, S-Down slightly increases inference time due to the limited number of nodes in SGraph, where the computational overhead introduced by sampling outweighs the benefits of node reduction. In contrast, D-Down reduces inference time, as DGraph contains significantly more nodes, and down-sampling leads to more efficient computation. Further comparisons show that the positive effects of sampling persist when SG and DG are combined, with no observed degradation in performance across other configurations.

\noindent\textbf{Effectiveness of the Information Fusion:}

\begin{itemize}
    \item (SG+SS) + (DG+PS) vs. (SG+SS) + (DG+PS) + IF
\end{itemize}

The comparison demonstrates that incorporating the information fusion module further enhances accuracy, accompanied by a slight increase in both the number of parameters and inference time. These results validate the effectiveness of the information fusion module in improving feature extraction capability.

\begin{table}[htbp]
\caption{Ablation studies of SDGraph architecture. Infer time: ms / instance.}
\centering
\begin{tabularx}{1.0\linewidth}{
    >{\raggedright\arraybackslash}m{0.02\linewidth} 
    >{\raggedright\arraybackslash}m{0.38\linewidth} 
    >{\centering\arraybackslash}m{0.11\linewidth} 
    >{\centering\arraybackslash}m{0.11\linewidth} 
    >{\centering\arraybackslash}m{0.15\linewidth} 
}

\toprule
\textbf{No.} & \textbf{Module} & \textbf{Acc.} & \textbf{Param.} & \textbf{Infer Time} \\

\midrule
1 & \textcolor{white}{(}SG & 0.6238 & 3,730,321 & 0.049 \\
2 & (SG + SS) & 0.6730 & 4,988,593 & 0.099 \\
3 & \textcolor{white}{(}DG & 0.6497 & 1,681,654 & 0.250 \\
4 & (DG + PS) & 0.6822 & 2,519,158 & 0.211 \\
5 & \textcolor{white}{(}SG + DG & 0.6522 & 5,593,607 & 0.272 \\
6 & (SG + SS) + DG & 0.6894 & 6,851,879 & 0.297 \\
7 & \textcolor{white}{(}SG + (DG + PS) & 0.7066 & 6,431,111 & 0.222 \\
8 & (SG + SS) + (DG + PS) & 0.7273 & 7,689,383 & 0.243 \\
9 & (SG + SS) + (DG + PS) + IF & 0.7537 & 8,411,063 & 0.301 \\

\bottomrule
\end{tabularx}
\label{ablation}
\end{table}

\section{Conclusions}

This paper introduced the Multi-Level Sketch Representation Scheme to identify the effective information for sketch representation learning, and proposed the SDGraph to fully leverage all the identified effective information. To validate the scheme, we conducted experiments by either excluding the effective information or incorporating not considered information. In both cases, performance consistently declined, confirming the validity of our conclusions. We further evaluated the performance of SDGraph on five tasks: sketch classification, CL-ZS-SBIR, FG-ZS-SBIR, FG-SBIR, and sketch generation. Experimental results demonstrate that SDGraph outperforms the state-of-the-art across all tasks. To assess the architectural design, we performed ablation studies focusing on the contributions of the SGraph, DGraph, graph node sampling, and information fusion module. The results confirm that each component contributes to performance improvement.

Despite the above effectiveness, this work has two limitations. First, although we analyze sketch information at the sketch-, stroke-, and point-levels, which is sufficient for sketch representation partitioning, the information considered within each level may not be exhaustive. Second, due to the varying number and length of strokes across sketches, we employ padding to standardize the input format. However, this approach increases storage demands and inference latency. Addressing these limitations will be the focus of our future work.

\bibliographystyle{IEEEtran}
\bibliography{reference}

\end{document}